\documentclass[floatfix,twocolumn,preprintnumbers,amsmath,amssymb,superscriptaddress]{revtex4-2}
\usepackage{amsmath,amssymb,eucal,graphicx,float,epstopdf,xparse}
\usepackage{epsfig,subfigure}
\usepackage[utf8]{inputenc}
\usepackage{comment}

\def \equi#1{\mathrel{\mathop{\kern 0pt\sim}\limits_{#1}}} 
\usepackage{xcolor}

\newcommand{\dw}{d_\text{w}}
\usepackage[shortlabels]{enumitem}

\begin{document}

\title{Record Ages of non-Markovian Scale-Invariant Random Walks}

\author{L\'eo R\'egnier}
\address{Laboratoire de Physique Th\'eorique de la Mati\`ere Condens\'ee,
CNRS/Sorbonne Universit\'e, 4 Place Jussieu, 75005 Paris, France}
\author{Maxim Dolgushev}
\address{Laboratoire de Physique Th\'eorique de la Mati\`ere Condens\'ee,
CNRS/Sorbonne Universit\'e, 4 Place Jussieu, 75005 Paris, France}
%\address{Laboratory of Theoretical  Physics and Condensed Matter (LPTMC),
%CNRS/Sorbonne University, 4 Place Jussieu, 75005 Paris, France}

\author{Olivier B\'enichou}
\email{benichou@lptmc.jussieu.fr}
\address{Laboratoire de Physique Th\'eorique de la Mati\`ere Condens\'ee,
CNRS/Sorbonne Universit\'e, 4 Place Jussieu, 75005 Paris, France}

\begin{abstract}
\begin{center}
    {\bf ABSTRACT}
\end{center}
How long is needed for an observable to exceed its previous highest value and establish a new record? This  time, known as the age of a record plays a crucial role in quantifying record statistics. Until now, general methods for determining  record age statistics have  been limited to observations of either independent  random variables or successive positions of a  Markovian (memoryless) random walk. Here we develop a theoretical framework to determine record age statistics in the presence of  memory effects for continuous non-smooth processes that are asymptotically scale-invariant.  Our theoretical predictions are confirmed by numerical simulations and experimental realizations of diverse representative non-Markovian random walk models and real time series with memory effects, in fields as diverse as genomics, climatology, hydrology, geology and computer science. Our results reveal the crucial role of the number of records already achieved in time series and  change our view on analysing record  statistics.

\end{abstract}
\maketitle

\section*{Introduction}

\begin{figure}[th!]
    \includegraphics[width=\columnwidth]{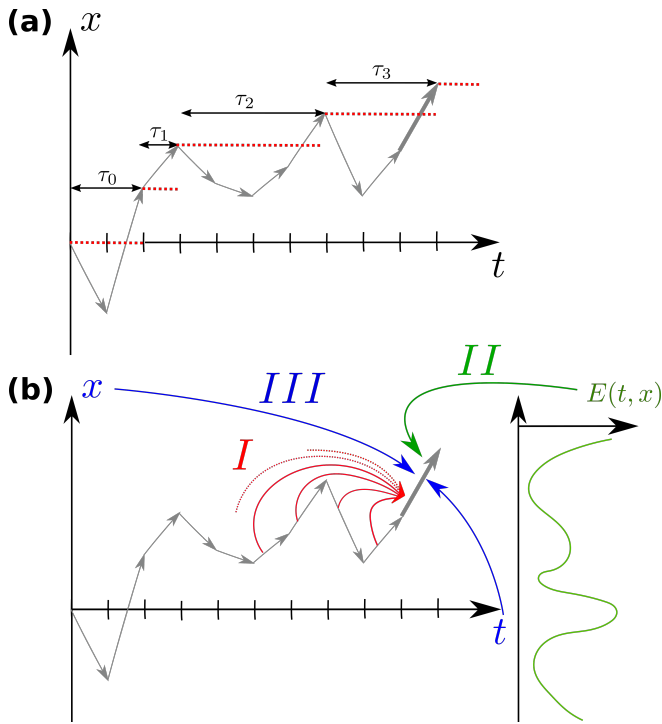}
    \caption{{\bf Record ages for non-Markovian random walks (RWs).} {\bf (a)} Sketch of a space time trajectory of the RW represented by successive discrete steps $\eta_t$ (grey arrows). The records in the trajectory are identified  by  red dotted lines. The record age $\tau_n$ of the RW  is defined as the time between the $n^\text{th}$ and $(n+1)^\text{st}$ records. {\bf (b)} Different statistical mechanisms giving rise to a non-Markovian
evolution: The statistics of the RW steps $\eta_t$ may depend on ($I$)  the previous steps of the walk (red arrow), ($II$) the environment with which the RW interacts (green arrow, schematically represented by the function $E(t,x)$), or ($III$) the current time or position (blue arrows). In this article, we show  that these memory effects strongly modify the record age statistics, which are no longer simply given by
the usual persistence exponent $\theta$, but also by a distinct exponent that we determine explicitly.}
    \label{fig:Illustration}
\end{figure}

The statistics of records in a discrete time series $\left( X_t \right)_{t =0,1,\ldots }$ is one of the main topics of interest in the study of extreme events \cite{majumdar2020extreme}, with applications in an increasing number of fields.
A record event occurs at time $t$ if all prior observations $\left( X_{t'} \right)_{t' =0,\ldots,t-1}$  are smaller than the last value $X_t$. In this context, the inter record times $\tau_n$, also called  record ages \cite{Kearney_2020,Godreche2021,godreche2017record,Kumar2023universal,Sabhapandit_2011,Benigni_2018,lacroix2020universal,aliakbari2017records},   between the $n^\text{th}$ and $(n+1)^\text{st}$ record, are pivotal, as they characterize the time of occurrence of the next record breaking event such as heatwaves \cite{morit2022}, earthquakes \cite{ambraseys1971,Ben2013} or record temperatures \cite{coumou2013}.

The theory of records has been studied since the mid-20th century \cite{chandler1952,nevzorov1988}, and is well understood when the random variables $\left( X_t \right)_{t =0,1,\ldots }$ are independent and identically distributed (i.i.d.) \cite{Eliazar2009,Krug_2007,gouet2020exact}.
An important step in the study of records was recently made when observations are  the successive positions of a Markovian RW \cite{godreche2017record,majumdar2008,majumdar2012record,godreche2014universal,Ben-Naim2014}, $X_{t+1}=X_t+\eta_{t+1}$, where the steps $\left( \eta_t \right)_{t =0,1,\ldots }$ are still i.i.d. and symmetric. In this situation, record ages are strictly given by the time $T$ needed to reach a given value for the first time, regardless of the past. This time follows an   algebraic tail distribution $\mathbb{P}(T \geq \tau)\propto \tau^{-\theta}$, where $\theta$ is the persistence exponent \cite{bray2013persistence}, provided by the celebrated Sparre-Andersen theorem \cite{klafter2011first}, yielding $\theta=1/2$. 
We emphasize that, despite the fact that this RW model accounts for correlations between the observations $\left( X_t \right)_{t =0,1,\ldots }$, the steps $\left( \eta_t \right)_{t =0,1,\ldots }$ themselves are independent.  As a result, this model cannot account for memory effects in the increments.

However, as a general rule, real time series are not only correlated but also exhibit such memory effects.  When the evolution  of an observable is influenced by interactions with hidden degrees of freedom, such as  the previous steps of the RW or its interaction with the environment, it cannot be  modeled as a Markov process. 

This is typically the case for displacement data from various  tracers (microspheres, polymers, cells, vacuoles...) in simple \cite{franosch2011resonances} and viscoelastic fluids \cite{Krapf2019,Weiss2013,Reverey2015}, soil  \cite{Crescenzo2022,sabbarese2020} and air temperatures \cite{brody2002}, river flows \cite{zhang2008,movahed2008}, nucleotide sequence locations \cite{peng1992long,Peng1994} and Ethernet traffic \cite{leland1991high,Fowler1991,leland1993}.
So far, as highlighted in the recent review \cite{godreche2017record}, almost nothing is known about the  record age statistics of non-Markovian processes.   The only exceptions concern processes   amenable to a Markovian process by adding an extra degree of freedom \cite{lacroix2020universal,Godreche2021,Gabel2012}, and a numerical observation in the specific case of the fractional Brownian motion \cite{aliakbari2017records}. Here, we provide a general scaling theory which determines the  time dependence of the record age statistics of non-Markovian RWs. We show  that memory effects significantly alter these statistics. They  are no longer solely governed by the persistence exponent $\theta$, but  also  by another   explicitly calculated exponent, which is the hallmark of non-Markovian dynamics. 

\section*{Results}
\subsection{Main Results}
We consider a general non-Markovian symmetric RW, whose successive positions  form a time series $\left( X_t \right)_{t =0,1,\ldots }$.  These positions satisfy $X_{t+1}=X_t+\eta_{t+1}$, where now the statistics of the steps  $\left( \eta_t \right)_{t =0,1,\ldots }$    may  exhibit ($I$) long-range correlations, ($II$) interactions with the environment (e.g. footprints left along the trajectory), or ($III$) explicit space-time dependence (see Fig. \ref{fig:Illustration}). Essentially all statistical mechanisms that lead to  non-Markovian evolution  are encompassed by these features of $X_t$  \cite{BOUCHAUD1990}. In turn, they allow to account for a variety of real time series displaying memory effects \cite{Magdziarz2009,mandelbrot1968fractional}.
At large time,  $X_{t}$  is assumed to converge  to a scale-invariant process that is continuous (i.e., excluding  broadly distributed steps $\eta_t$) and non-smooth \cite{bray2013persistence} (meaning that, as for the standard Brownian motion,  the trajectory is irregular, having at each point an infinite derivative). Under these conditions, the process is characterized by a walk dimension \cite{BOUCHAUD1990} $\dw>1$, such that $X_t \propto t^{1/\dw}$, and  the random variable $X_t/t^{1/\dw}$ is asymptotically independent of $t$. To account for  potential aging in the increments,  $X_t$ is more generally assumed to have scale-invariant increments,  meaning that, for  $1\ll t \ll T$, $X_{t+T}-X_T \propto t^{1/\dw^0}T^{ \alpha/2}$.  This  defines the aging exponent  $\alpha$ \cite{Schulz2014,Levernier2018} ($\alpha>0$ corresponding qualitatively to accelerating
processes and $\alpha<0$ to slowing down processes) and  an effective walk dimension at short times  $\dw^0\equiv (\dw^{-1}-\alpha/2)^{-1}$. We stress that  the class of processes that we consider here covers a very broad range of examples  of non-Markovian RWs, as detailed below, despite not covering the particular cases of L\'evy flights \cite{majumdar2008} (which are discontinuous) or of the Random Acceleration Process \cite{Godreche2021} (smooth), which would require a different approach.

We report   that  the  tail distribution ${S(n,\tau)\equiv \mathbb{P}(\tau_n \geq \tau)}$ of the record age $\tau_{n}$ asymptotically obeys a  scaling behaviour $S(n,\tau)=n^{-1}\psi(\tau/n^{\dw})$, displaying two universal distinct algebraic regimes : 

\begin{equation}
S\left(n,\tau\right)  \propto \left\{
\begin{array}{ll}
       \frac{1}{n}\left(\frac{n^{\dw}}{\tau}\right)^{\frac{1}{\dw^0}}  &\mbox{for }  \; n^{\dw-\dw^0}  \ll \tau \ll n^{\dw } ,\\[\bigskipamount]
         \frac{1}{n}\left(\frac{n^{\dw}}{\tau}\right)^{\theta} &\mbox{for }  \; 1 \ll n^{\dw} \ll \tau \; ,\\
    \end{array}
    \right.
\label{eq:ScalingResultRecords}
\end{equation}
where $\psi$ is a process dependent scaling function and the persistence exponent $\theta$ has been defined above. 
Equation \eqref{eq:ScalingResultRecords} explicitly determines the $n$ and $\tau$ dependence of  the record age statistics of non-Markovian RWs. 
Fundamental consequences of our results include:
(i) In regime 1, defined by  $n^{\dw-\dw^0}  \ll \tau \ll n^{\dw }$, the record time's decay is governed by an exponent different from  $\theta$. While it is not unexpected that the memory of the past affects record age statistics for a non-Markovian process (in particular, it is known that it can change the persistence exponent \cite{majumdar1996,levernier2022}), it is striking that the corresponding exponent is fully explicit and depends only on the effective walk dimension $\dw^0$ of the increments. Note that regime 1 can span several orders of magnitude as soon as sufficiently many records have been broken, and thus dominate the observations. (ii) In regime 2, defined by $\tau\gg n^{\dw}$, the decay in the record time can be very different from that of regime 1. This is particularly striking for processes with stationary increments for which the exponent involved in regime 2, $\theta=1-1/\dw$ \cite{Levernier2018}, is markedly different from  the exponent $1/\dw^0=1/\dw$ of regime 1 (with the exception of Markovian RWs for which the two exponents are both $1/2$ and a single regime is recovered; note that this single regime of exponent 1/2 is also obtained in the case of Lévy flights, which are not covered by our approach).  (iii) The  record age distribution ages, in the sense that it depends on the number $n$ of records  already achieved. 
Consequently,  the observations of early record ages are not representative of  later records and call for a careful analysis of real data (note that the record distribution also ages in  time series with i.i.d. observations $X_t$, which are thus not of the form $X_{t+1}=X_t+\eta_{t+1}$ considered here, but the dependence of this distribution on the number of records and the corresponding statistical mechanisms are very different \cite{godreche2017record}).
Finally, note that despite the existence of  two regimes  for record ages, because of the explicit dependence of  the prefactors of $S(n,\tau)$ on $n$, the number of records at time $t$ displays a single time regime $n\propto t^{1/\dw}$ (see Supplementary Information, SI). 

\subsection{Derivation of the results}

The following outlines the derivation of these results   (see SI Sec.~S1 for details):

The first step consists in noting that, due to the scale-invariance of the process $X_t$, the time $T_n$ to reach the $n^\text{th}$ record, $T_n\equiv \sum_{k=0}^{n-1}\tau_k$, satisfies  $T_n\propto n^{\dw}$ and its increments obey $T_{m+n}-T_m\propto m^{\dw-\dw^0}n^{\dw^0}$ (see SI Sec.~ S1.B). In other words, $\mathbb{P}\left(T_{m+n}-T_m \leq T \right)$ is a function of a single variable $T/(m^{\dw-\dw^0}n^{\dw^0})$. Then,  $T_{m+n}-T_m=\sum_{k=m}^{n+m-1} \tau_k$ is dominated by the largest record age \cite{BOUCHAUD1990,Vezzani2019} under the self-consistent assumption that $S(n,\tau)\propto n^{-1+\epsilon_1}\tau^{-y_1}$ for $\tau \ll n^{\dw}$ (regime 1) and $S(n,\tau)\propto n^{-1+\epsilon_2}\tau^{-y_2}$ for $\tau \gg n^{\dw}$ (regime 2) with $y_i$ between 0 and 1. This results in the equation 
\begin{align}
\label{Eqstart}
    \mathbb{P}(T_{m+n}-T_m \leq T) \simeq \mathbb{P}(\max(\tau_m,\ldots, \tau_{m+n-1}) \leq T).
\end{align}
Adapting the argument of Ref.~\cite{carpentier2001}, we show for continuous scale-invariant non-smooth processes analytically (see Sec.~S1.D of SI) and verify numerically (see Sec.~S2.C of SI) that, in Eq.\eqref{Eqstart}, the record ages  $\tau_k$  are asymptotically ($n \gg 1$) effectively independent, which leads  to  

\begin{align}
    \mathbb{P}(T_{m+n}-T_m \leq T) \simeq \prod_{k=m}^{n+m-1} (1-S(k,T)) \; .
    \label{eq:indepApprox}
\end{align}
First, for time scales $T$ much smaller than the typical time $T_m\propto m^{\dw}$ required to  break $m$ records and for $n \ll m$ (regime 1),  Eq.~\eqref{eq:indepApprox} becomes
\begin{align}
    \mathbb{P}(T_{m+n}-T_m \leq T) \underset{T,n^{\dw} \ll m^{\dw}}{\propto} \exp\left[-\frac{\text{const. } n}{m^{1-\epsilon_1}T^{y_1}} \right] \; .
\end{align}
Using  $T_{m+n}-T_m\propto m^{\dw-\dw^0}n^{\dw^0}$ gives the  exponents of regime 1 as $y_1=1/\dw^0$ and $\epsilon_1=\dw/\dw^0$. Second, for $\tau \gg n^{\dw}$ (regime 2), the memory of the $n$ broken records no longer affects the algebraic time decay of $S(n,\tau)$, which is thus given by the persistence exponent $\theta=y_2$. Taking $m=0$ in Eq.~\eqref{eq:indepApprox}, we get
\begin{align}
    \mathbb{P}(T_n\leq T) \propto \exp\left[-\text{const. } n^{\epsilon_2}/T^\theta \right].
\end{align}
Using  $T_n\propto n^{\dw}$ leads to the exponent $\epsilon_2=\dw \theta$.

\subsection{Comparison with numerical simulations of non-Markovian models}

\begin{figure*}[th!]
    \centering
    \includegraphics[width=2\columnwidth]{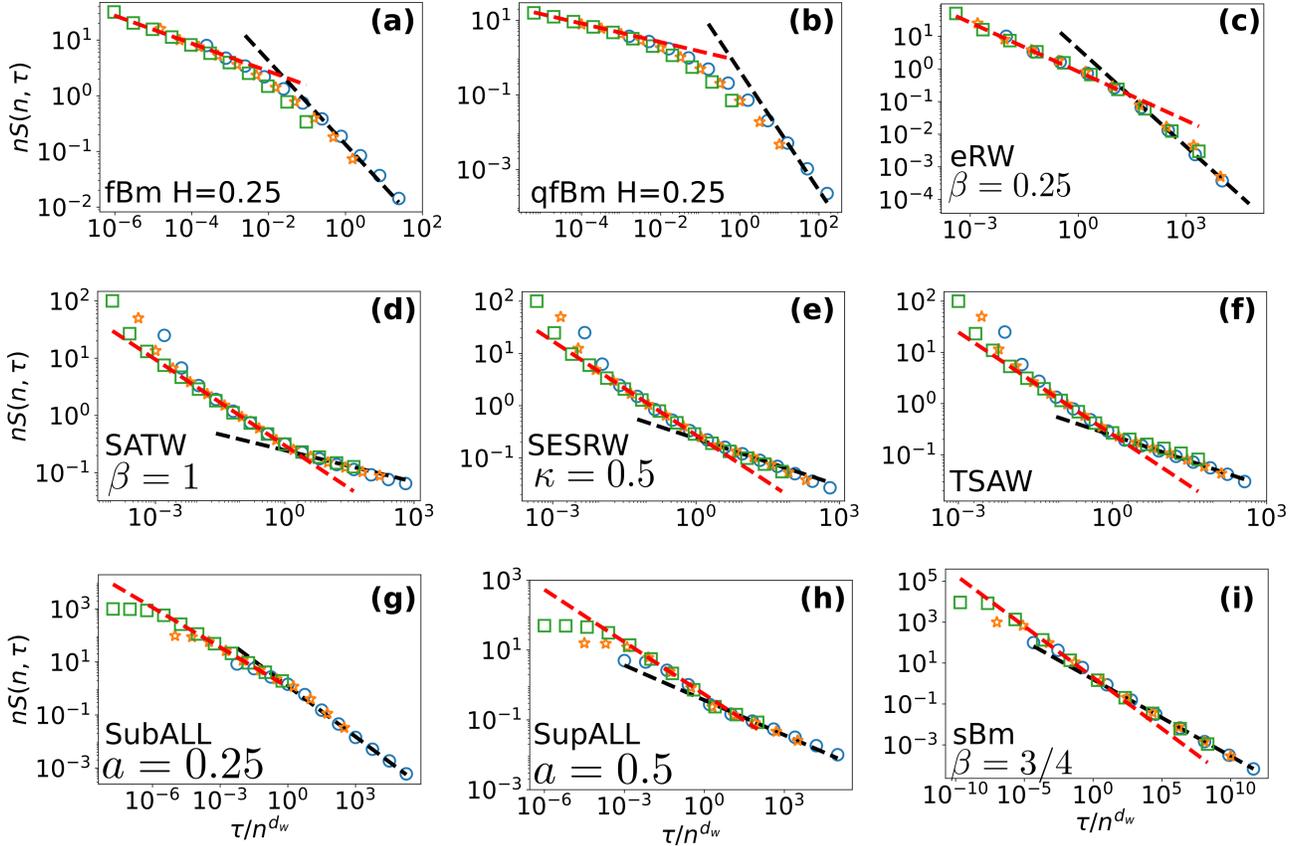}
    \caption{{\bf Universal record age distributions for non-Markovian RWs: theoretical predictions (lines) vs numerical simulations (symbols).} Simulated rescaled tail distribution of record ages $\tau_n$ for different values of record number $n$ displayed for various representative RW models: \textbf{(a)} fractional Brownian motion (fBm) of Hurst exponent $H=0.25=1/\dw=1-\theta$ for $n=8,$ $16$ and $32$ \textbf{(b)} quenched fBm (qfBm) of Hurst exponent $H=0.25=1/\dw$ and $\theta\approx 1.55$ for $n=5$, $10$ and $20$ \textbf{(c)} elephant RW (eRW) with $\beta=0.25$ such that $\dw=2$ and $\theta=1$, for $n=10$, $25$ and $50$ \textbf{(d)} Self-Attractive Walk (SATW) with $\beta=1$, such that $\dw=2$ and $\theta=e^{-1}/2$ for $n=25$, $50$ and $100$ \textbf{(e)} Sub-Exponential Self-Repelling Walk (SESRW) with $\beta=1$ and $\kappa=0.5$ such that $\dw=5/3$ and $\theta\approx 0.3$ for $n=25$, $50$ and $100$ \textbf{(f)} True Self-Avoiding Walk (TSAW) with $\beta=1$ such that $\dw=3/2$ and $\theta=1/3$ for $n=25$, $50$ and $100$ \textbf{(g)} subdiffusive Average L\'evy Lorentz (subALL) with $a=0.25$ such that $\dw=2.75$, $\dw^0=2$ and $\theta=7/11$ for $n=10$, $100$ and $1000$ \textbf{(h)} superdiffusive ALL (supALL) with $a=0.5$ such that $\dw=3/2$, $\dw^0=2$ and $\theta=1/3$ for $n=10,$ $100$ and $1000$ \textbf{(i)} exact rescaled tail distribution (see SI) for scaled Brownian motion (sBm) with $\beta=0.75$ such that $\dw=8/3$, $\dw^0=2$ and $\theta=3/8$ for $n=100$, $1000$ and $10000$.  Increasing values of $n$ are represented respectively by blue circles, orange stars and green squares. The black dashed line represents the algebraic decay $\tau^{-\theta}$ while the red dashed line stands for the algebraic decay $\tau^{-1/\dw^0}$.}
    \label{fig:All}
\end{figure*}

We  confirm the validity of our analytical results in Fig.~\ref{fig:All} by comparing them to numerical simulations of a broad range of representative RW examples, which illustrate the  classes ($I$), ($II$), and ($III$) of non-Markovianity discussed above.  
Specifically, we consider  (see SI for precise definitions and Supplementary Table 1 for a summary of their characteristics): ($I$)  (a) the fractional Brownian motion (fBm), a non-Markovian  Gaussian process, with stationary increments given by $\langle (X_t - X_0)^2\rangle = t^{2H}$, where $H$ is the Hurst exponent;   this paradigmatic model has been used repeatedly to account for anomalous diffusion induced by long-range correlations in viscoelastic fluids \cite{Krapf2019} as well as temporal series displaying memory effects \cite{mandelbrot1968fractional,Magdziarz2009}; (b)  its extension to quenched initial conditions (qfBm), for which the statistics of increments is not stationary anymore,  and which describes for instance the height fluctuations under Gaussian noise of an initially flat interface \cite{majumdar1996,Levernier2018}; 
(c) the elephant RW (eRW) \cite{Schutz2004}, for which the current step is drawn uniformly from all of the previous steps performed by the RW, and then reversed with probability $\beta$; ($II$) (d) The Self-Attractive Walk (SATW), (e) Sub-Exponential Self-Repelling Walk (SESRW) and (f) True Self-Avoiding Walk (TSAW) are prototypical examples of self-interacting RWs \cite  {sapozhnikov1994self,davis_1990,barbier2020anomalous,barbier2022self},   for which the RW deposits a signal at each lattice site it visits and then has a transition probability depending on the number of visits to its neighbouring sites (see SI for precise rules), so that memory emerges from the interaction of the walker with the territory already visited; these RWs have been shown to be relevant in the case of living cells, where it was demonstrated experimentally that  various cell types can chemically modify the extracellular matrix, which in turn deeply impact their motility \cite{Alessandro2021}; ($III$) Two models involving  an explicit  spatial or temporal dependence of  the steps:  (g) the subdiffusive (resp. (h) the superdiffusive) Average L\'evy Lorentz model (subALL and supALL, respectively) \cite{Radice_2020,Radice2020_2,Barthelemy2008} for which the transmission (resp. reflection) probability at every site decays algebraically with the distance to the origin, and (i) the scaled Brownian motion (sBm) \cite{Lim2002} for which the jumping rate is an algebraic function of time, and which is a paradigmatic model of subdiffusion \cite{saxton2001anomalous}.

Figure \ref{fig:All} reveals excellent quantitative agreement between numerical simulations and our analytical results. The data collapse of the properly rescaled record ages tail distribution and the confirmation of the two successive algebraic decays $\tau^{-1/\dw^0}$ and $\tau^{-\theta}$ show that  Eq.~\eqref{eq:ScalingResultRecords}  unambiguously captures the dependence on both the number of records $n$ and the time $\tau$ (further confirmed by the analytical determination of the full tail distribution  in the solvable case of the  sBm, see SI).  We emphasize that the very different nature of these examples (subdiffusive and superdiffusive, aging and non aging, covering all classes of  non-Markovian RWs) shows the broad   applicability of our approach.

\section*{Discussion} 

\begin{figure*}[th!]
    \centering
    \includegraphics[width=2.1\columnwidth]{Data.pdf}
    \caption{ {\bf Universal record age distributions for non-Markovian RWs: theoretical predictions (lines) vs experimental RW realizations and real time   observations (symbols).}\\ 
    {\bf (a)}-{\bf (h)} Distribution of the increment $x_t=X_{t+T}-X_{T}$ at different times $t$ normalised by $t^{1/\dw}$ for: 
    %where $\dw$ was obtained via the DMA method   
    {\bf (a)} river discharge ($t=10$, $20$, and $40$), {\bf (b)} volcanic soil temperature ($t=5$, $10$, and $20$), {\bf (c)} motion of microspheres in a gel ($t=2$, $4$, and $8$), {\bf (d)} motion of vacuoles inside an amoeba ($t=10,$ $20$, and $40$), {\bf (e)} motion of telomeres ($t=20$, $40$, and $80$),  {\bf (f)} DNA RW ($t=20$, $40$, and $80$) , {\bf (g)} cumulative air temperature ($t=5$, $10$, and $20$), and  {\bf (h)} Ethernet cumulative requests ($t=500$, $1000$, and $2000$). Increasing values of times are represented successively by blue circles, orange stars and green squares.\\
    {\bf (a$^\prime$)}-{\bf (d$^\prime$)} Statistics of the time to first reach the initial value in the sub interval (blue stars) and the statistics of the records (regardless of the number $n$ of records, orange circles) for {\bf (a$^\prime$)} river discharge, {\bf (b$^\prime$)} volcanic soil temperature, {\bf (c$^\prime$)} motion of microspheres in a gel, and {\bf (d$^\prime$)} motion of vacuoles inside an amoeba. The black dashed line represents the algebraic decay $\tau^{-1+1/\dw}$ while the red dashed line stands for the algebraic decay $\tau^{-1/\dw}$. \\
    {\bf (e$^\prime$)}-{\bf (h$^\prime$)} Rescaled tail distribution of record ages $\tau_n$  for different values of the number of records $n$ for {\bf (e$^\prime$)} motion of telomeres ($n=1$, $3$, and $6$), {\bf (f$^\prime$)} DNA RW ($n=1,$ $2$, and $4$), {\bf (g$^\prime$)}  cumulative air temperatures ($n=1$, $2$, and $3$), and {\bf (h$^\prime$)} Ethernet cumulative requests ($n=1$, $5$, and $25$). Increasing values of $n$ are represented successively by blue circles, orange stars, and green squares. The lines represent the algebraic decays as for {\bf (a$^\prime$)}-{\bf (d$^\prime$)}.\\}
    \label{fig:Data}
\end{figure*}

We demonstrate the relevance  of our results by showing that they apply even when the hidden degrees of freedom responsible for the non-Markovianity of the dynamics are unknown, as is the rule in real observations.

This is illustrated by considering both   trajectories  involving  a variety of tracers in complex fluids (see Fig.\ref{fig:Data}  {\bf (c)} to {\bf (e)}, which provide experimental realizations  \cite{Krapf2019} of several non-Markovian RW models discussed above) and  real time series in  diverse  fields displaying memory effects, for which record ages are crucial as they characterize  the occurrence of extreme events (see Fig.\ref{fig:Data}  {\bf (a)}, {\bf (b)} and {\bf (f)} to {\bf (h)}). 

Specifically, we consider the following data: {\bf (a)} river flows \cite{zhang2008} ($1/\dw \approx 0.14$), {\bf (b)} volcanic soil temperatures \cite{Crescenzo2022,sabbarese2020} ($1/\dw \approx 0.42$),  {\bf (c)} trajectories of microspheres  in gels \cite{Krapf2019} ($1/\dw \approx 0.43$) {\bf (d)} trajectories of vacuoles  inside an amoeba \cite{Krapf2019} ($1/\dw \approx 0.67$), {\bf (e)} trajectories of telomeres in a nucleus \cite{Krapf2019,stadler2017} ($1/\dw \approx 0.25$), {\bf (f)} pyrimidines/purines DNA RW  where a step value is given by the nucleotide type, $+1$ for adenine/thymine, $-1$ for cytosine/guanine \cite{Peng1994,peng1992long} ($1/\dw \approx 0.67$), {\bf (g)} cumulative air temperatures \cite{brody2002} ($1/\dw\approx 0.8$), {\bf (h)} cumulative Ethernet traffic \cite{leland1991high,Fowler1991,leland1993} ($1/\dw \approx 0.8$). The walk dimension $\dw$  was estimated by applying  the Detrending Moving Average (DMA) method \cite{holl2019,alessio2002second} to these data, which removed the  deterministic behaviours  (see SI for details on the datasets' analysis). 
Indeed, the characterization of extreme events, and thus records, requires the meticulous examination of fluctuations around the trend, as underlined in \cite{brody2002,amaya2023marine}.

We stress that we do not require any knowledge on the microscopic details of the process to obtain the record age statistics provided by Eq.~\eqref{eq:ScalingResultRecords}. In particular, the processes are not necessarily Gaussian and can exhibit various distributions of the increments $x_t \equiv X_{T+t}-X_T$ (see Fig.~\ref{fig:Data}), as long as they are asymptotically scale-invariant (the sampling time of the data is much longer than the microscopic time scales involved in the process to avoid effects similar to those observed in \cite{Eli2022}, as it is checked in Sec.~S3 of SI).

Figure \ref{fig:Data} demonstrates  the quantitative agreement between various  real data (see SI Supplementary Figure 8 for additional datasets, including examples displaying aging of the increments $x_t$) and our analytical predictions given by Eq.~\eqref{eq:ScalingResultRecords}. The strong dependence of record ages  on the number $n$ of records already achieved, predicted by our analytical approach and confirmed  by both numerical simulations and real observations, is a direct manifestation of the non-Markovian feature of the underlying RWs. These results quantitatively demonstrate the significance of memory effects in the record ages of  non-Markovian RWs,  providing the tools to better  predict   record-breaking events.

\section*{Methods}
\subsubsection*{Numerical simulations of non-Markovian RWs}
In this section, we present briefly the models and the numerical methods used to generate the data in Fig \ref{fig:All}.
\begin{itemize}
\item[({\bf a})] \textit{Fractional Brownian motion (fBm).} The fBm is a non-Markovian Gaussian process, with stationary increments. Thus, an fBm $X_t$ of Hurst index $H$ is defined by its covariance 
\begin{align}
    \text{Cov}\left(X_t,X_{t'} \right)= \frac{1}{2} \left(t^{2H} + t'^{2H}-|t-t'|^{2H} \right) \; .
\end{align}
The steps $\eta_t=X_t-X_{t-1}$ are called fractional Gaussian noise (fGn). Nowadays, the fBm is broadly spread and its implementations could be found in standard packages of python or Wolfram Mathematica.
\item[({\bf b})]\textit{Quenched fBm (qfBm).} This process is an extension of fBm to quenched initial conditions, which results in non-stationary increment statistics. In particular, it describes the height fluctuations under Gaussian noise of an initially flat interface. Then $X_t$ corresponds to the height of the interface at position $x=0$, $X_t=h(0,t)$, $h(x,t)$ following the Stochastic Differential Equation (SDE)
\begin{align}
    \partial_t h(x,t)=-\left(-\Delta \right)^{z/2} h(x,t) + \eta(x,t).
\end{align}
Here $\eta(x,t)$ is a Gaussian noise with possible spatial correlations. We solve numerically this SDE with a spatial discretization $\Delta x=1$ and a time discretization $\Delta t=0.1$. The system is initially flat, $h(x,t=0)=0$. 
\item[({\bf c})]\textit{Elephant RW (eRW).}
This process is representative of interactions with its own trajectory. At time $t$, the step $\eta_t$ is drawn uniformly among all the previous steps $\eta_i$ ($i<t$) and is reversed with probability $\beta$. 
\item[({\bf d})]\textit{Self-attractive walk (SATW).} This model is a
prototypical example of self-interacting RWs. In the SATW model \cite{sapozhnikov1994self,davis_1990,barbier2020anomalous,barbier2022self}, the RW at position $i$ jumps to a neighbouring site $j=i \pm 1$ with probability depending on the number of times $n_j$ it has visited site $j$, 
\begin{align}
    p(i\to j)=\frac{\exp \left[ -\beta H(n_j) \right]}{\exp \left[ -\beta H(n_{i-1}) \right]+\exp \left[ -\beta H(n_{i+1}) \right]},
\end{align}
where $H(0)=0$, $H(n>0)=1$ and $\beta>0$..
\item[({\bf e-f})]\textit{Exponential self-repelling RW.} This is another example of self-interacting RW. In this model, the RW at position $i$ jumps to a neighbouring site $j=i \pm 1$ with probability depending on the number of times $n_j$ it has visited site $j$, 
\begin{align}
    p(i\to j)=\frac{\exp \left[ -\beta n_j^\kappa \right]}{\exp \left[ -\beta n_{i-1}^\kappa \right]+\exp \left[ -\beta n_{i+1}^\kappa \right]}
\end{align}
where $\kappa$ and $\beta$ are two positive real numbers.
\item[({\bf g-h})]\textit{Average L\'evy Lorentz gas (ALL).} We consider a RW on a $1d$ lattice with position dependent reflection or transmission probabilities $r(x)$ or $t(x)$. In the subdiffusive model (resp. superdiffusive model), the transmission coefficient $t(x)$ (resp. reflection coefficient $r(x)$) is taken to be proportional to $|x|^{a-1}$ at large distance $|x|$ from the origin.
\end{itemize}

\subsubsection*{Data analysis}
In this section we provide the method developed to determine the walk dimension of the time series presented in Fig.~\ref{fig:Data} as well as numerical checks of their stationarity.\\

{\it(i)  Walk dimension determination: }In order to obtain the walk dimension $\dw$ in a time series, we apply the Detrending Moving Average (DMA) method~\cite{alessio2002second,holl2019}, which consists in evaluating the typical fluctuations in a window of size $\ell$ regardless of any bias or deterministic trend. More precisely, for a dataset $(X_t)_{t=0,\ldots,N}$, we consider the windows of size up to $\ell_\text{max}$, compute the window averages $x_t^\ell=\frac{1}{\ell}\sum_{i=0}^{\ell-1}X_{t-i}$, and the typical fluctuation for a window of size $\ell$, $F(\ell)=\sqrt{\frac{1}{N-\ell_\text{max}}\sum_{t=\ell_\text{max}}^N (X_t-x_t^\ell)^2}$. When several trajectories are available, we consider the average fluctuation over all the trajectories (for telomeres, vacuoles and microspheres in agarose data). If the data behave as a RW of walk dimension $\dw$, then $F(\ell) \propto \ell^{1/\dw}$. We obtain the value of $1/\dw$ via the DMA method to each dataset.  \\
\\
{\it(ii) Check of stationarity: }In order to check that the data are stationary, we compare the MSD obtained from the increments $\lbrace x_t=X_{t+T}-X_T \rbrace_{T\leq N/4,t}$ in the first quarter of the data and the increments $\lbrace x_t=X_{t+T}-X_T \rbrace_{3N/4 \leq T,t}$ in the last quarter of the data.   \\
\\
{\it(iii) Record ages in datasets: }Record ages are obtained by starting the subtrajectories at values of $t$ equally spaced at intervals at least $200$ time steps long, and observing successive records occurring in the subtrajectory. First return times are obtained by starting the subtrajectories at any value of time.

\section*{Data availability}
 
 The simulation data of this study are generated based on the code deposited in a GitHub repository \cite{codeRecords} located at \url{https://github.com/LeoReg/RecordAges}.

 The data of the Hadley Centre Central England Temperature (HadCET) project are available at \url{https://www.metoffice.gov.uk/hadobs/hadcet/}. The data of the European Climate Assessment \& Dataset (ECA\&D) project are available at \url{https://www.ecad.eu/}. The volcanic soil temperature data are available at Ref.~\cite{sabbarese2020}. River discharge data are available at \url{https://portal.grdc.bafg.de/applications/}. The GenBank database is available at \url{https://www.ncbi.nlm.nih.gov/genbank/}. The data of traffic traces are available at \url{http://ita.ee.lbl.gov/html/contrib/BC.html}. Experimental trajectories of fBm realizations are available upon request by the authors of Ref.~\cite{Krapf2019}. Experimental cell migration trajectories are available upon request by the authors of Ref.~\cite{Alessandro2021}.

\section*{Code availability}
The codes used to generate the simulation data presented in this study as well as the code to analyze the experimental data have been deposited in a GitHub repository located at \url{https://github.com/LeoReg/RecordAges}.

%\bibliography{ref}
%\bibliographystyle{naturemag}

\section*{Acknowledgements}
We thank T. Guérin, N. Levernier, and G. Oshanin for helpful discussions, G. Page for  careful reading of the manuscript, and  S. Majumdar for mentioning the similarity between the record age statistics and the statistics of the times between visits of new sites. 
We are thankful to D. Krapf, M. Weiss, F. Taheri and C. Selhuber-Unkel for providing us the experimental trajectories of fBm realizations used in Ref. \cite{Krapf2019}.
We thank J. d'Alessandro for providing us the experimental cell migration trajectories analysed in Ref. \cite{Alessandro2021}. We acknowledge the data providers in the Hadley Centre Central England Temperature (HadCET) and European Climate Assessment \& Dataset (ECA\&D) projects. We thank the authors of Ref. \cite{sabbarese2020} for giving access to the volcanic soil temperature data.  We acknowledge the Global Runoff Data Centre (GRDC), 56068 Koblenz, Germany for providing the Elbe and Rhône rivers' water debit data. We acknowledge the data providers of the GenBank database, hosted by the National Library of Medicine, as well as Jaenicke T., Diederich K.W., Haas W., Schleich J., Lichter P., Pfordt M., Bach A. and Vosberg H.P.  who deposited the specific HUMBMYH7 sequence used in this study. We thank the authors of Ref. \cite{leland1991high} for the data of traffic traces.

\section*{Author Contributions}
O.B., L.R. and M.D. contributed to analytical calculations. L.R. and M.D. performed numerical simulations. All the authors wrote the manuscript. O.B. conceived the research

\section*{Competing Interests}
The authors declare no competing interests.

\end{document}

\title{SUPPLEMENTARY INFORMATION\\
Record Ages of Non-Markovian Scale-Invariant Random Walks }

\author{L. R\'egnier}
\author{M. Dolgushev}
\author{O. B\'enichou}
\maketitle

\tableofcontents

\section{Scaling theory}
We provide here the detailed calculations corresponding to the general scaling theory developed in the main text.

\subsection{Definitions}

We assume that the random walk (RW) $(X_t)_{t=0,\ldots}$  converges at large time to a continuous and non-smooth scale-invariant process. Under these conditions, the process is characterized by a walk dimension $\dw>1$, such that $X_t \propto t^{1/\dw}$, and the random variable $X_t/t^{1/\dw}$ is independent of $t$. 
We more generally assume that  $X_t$ has scale-invariant increments with a potential ageing,  meaning that, for  $1\ll t \ll T$, $X_{t+T}-X_T \propto t^{1/\dw^0}T^{ \alpha/2}$.  This  defines the aging exponent  $\alpha$ \cite{Schulz2014,Levernier2018} and  an effective walk dimension at short times  $\dw^0\equiv (\dw^{-1}-\alpha/2)^{-1}$. 

To describe the properties of the RWs, we will use their continuous limit, which requires to introduce a microscopic cut-off either in time or space (as done in \cite{Godreche2021}). 
Relying here on a spatial cut-off $\Delta x$,  we consider the time $T_{x_0}$ to first reach the level $x_0=n\Delta x$ starting from the origin, as the continuous counterpart of the time $T_n$ to break $n$ records.

\subsection{Scale-invariance of the time $T_n$ to break $n$ records and of its increments}

First, we check that the temporal scale-invariance of $X_t$ leads to the spatial scale-invariance of $T_{x_0}$.
Indeed, since $X_t/t^{1/\dw}$ is  asymptotically independent of time $t$,  $X_{T_{x_0}}/T_{x_0}^{1/\dw}=x_0/T_{x_0}^{1/\dw}$ is asymptotically independent of $x_0$ \leo{($X_{T_{x_0}}=x_0$ because of continuity)}. Consequently,  $T_{x_0}$ is scale-invariant in the sense that $T_{x_0}/x_0^{\dw}$ is independent of $x_0$. Taking $\Delta x=1$ and $x_0=n\Delta x=n$ results in the following form for the cumulative distribution:
\begin{align}
    \mathbb{P}(T_n \leq T)&=\Phi\left(T/n^{\dw} \right) \; ,\label{eq:si_T}
\end{align}
where the scaling function $\Phi$ is independent of $T$ and $n$ (see \ref{fig:Averages} for numerical check of the number of records $n$ scale-invariance with time $T$ for representative non-Markovian RWs).\\

Second, we show that the scale-invariance of the increments $X_{t+T}-X_T$, $X_{t+T}-X_T\propto t^{1/\dw-\alpha/2}T^{\alpha/2}=t^{1/\dw^0}T^{\alpha/2}$, implies the scale-invariance of $T_{x_0+x_1}-T_{x_0}$. Consider the random variable $\frac{X_{T+t}-X_T}{T^{\alpha/2}t^{1/\dw^0}}$, which is independent of $t$ and $T$ as long as $1 \ll t \ll T$. By replacing $T$ by $T_{x_0}$ and $t$ by $ T_{x_0+x_1}-T_{x_0}$, we find that
\begin{align*}
    \frac{X_{T_{x_0+x_1}}-X_{T_{x_0}}}{T_{x_0}^{\alpha/2}(T_{x_0+x_1}-T_{x_0})^{1/\dw^0}}=\frac{x_1}{T_{x_0}^{\alpha/2}(T_{x_0+x_1}-T_{x_0})^{1/\dw^0}},
\end{align*}
 is  asymptotically independent of $x_0$ and $x_1$ for $1 \ll x_1 \ll x_0$. Using that $T_{x_0}/x_0^{\dw}$ is independent of $x_0$, we finally obtain that
\begin{align*}
    \frac{T_{x_0+x_1}-T_{x_0}}{(x_1/x_0^{\dw \alpha/2})^{\dw^0}}=\frac{T_{x_0+x_1}-T_{x_0}}{x_1^{\dw^0} x_0^{\dw - \dw^0}}
\end{align*}
is independent of $x_0$ and $x_1$. In other words, we have the scale invariance of $T_{x_0+x_1}-T_{x_0}$, ${T_{x_0+x_1}-T_{x_0}}\propto{x_1^{\dw^0} x_0^{\dw - \dw^0}}$. 
Taking $\Delta x=1$, $x_0=m \Delta x$ and $x_1=n \Delta x$ implies the scaling $T_{m+n}-T_m \propto n^{\dw^0}m^{\dw-\dw^0}$. It means that the cumulative distribution in the limit $1 \ll n\ll m$ and $1 \ll T\ll m^{\dw}$ can be written as
\begin{align}
    \mathbb{P}(T_{m+n}-T_m \leq T) = \Psi\left(\frac{T}{n^{\dw^0}m^{\dw-\dw^0}}\right),\label{eq:si_incT}
\end{align}
where the scaling function $\Psi$  is independent of $T$, $n$, and $m$. This scale-invariance of the time increments is systematically checked numerically  for a number of  representative non-Markovian RWs in \ref{fig:ScaleInv}.

\subsection{Characteristic exponents of the record age distribution}
 We  derive the exponents governing the algebraic decay of the record age distribution $S(n,\tau)$ by elaborating the arguments sketched in the main text.

We note by $\tau_k$ the $k^ \text{th}$ record age or, in the continuum setting, the time to reach level $(k+1)\Delta x$ starting from the first arrival at level $k\Delta x$. $T_n$ being the time to first reach level $n\Delta x$, it is given by the sum of the  \mx{$\{\tau_k\}$}, 
\begin{align}
    T_n=\sum_{k=0}^{n-1} \tau_k \; . \label{eq:defTn}
\end{align}
We  make the self-consistent assumption that the tail distribution of $\tau_k$ is algebraic of exponent smaller than $1$. As a consequence,  the sum \eqref{eq:defTn} is controlled by the largest $\tau_k$ \cite{BOUCHAUD1990}, i.e. 
\begin{align}
    \sum_{k=0}^{n-1} \tau_k \sim \max \left( \tau_0, \ldots , \tau_{n-1}\right) \; . \label{eq:summax}
\end{align}
We now assume (\leo{see the next subsection for an analytical justification, as well as}  \ref{fig:IndepApprox} and \ref{fig:Records} for systematic numerical checks for a number of representative non-Markovian RWs) \leo{that the record ages $\tau_k$ involved in Eq.~\eqref{eq:summax}} are asymptotically ($n \gg 1$) effectively independent, which leads  to  
\begin{align}
    \mathbb{P}(T_n \leq T) &\simeq \prod_{k=0}^{n-1} \mathbb{P}(\tau_k \leq T)= \prod_{k=0}^{n-1} (1-S(k,T)).\label{eq:dist_prod}
\end{align}

We search now the exponents $y_i$ and $\epsilon_i$ characterizing the tail distribution of $\tau_k$, $S(k,T)\propto k^{-1+\epsilon_1}T^{-y_1}$ for $ T \ll k^{\dw } $ (regime 1) and $S(k,T) \propto k^{-1+\epsilon_2}T^{-y_2} $ for $T \gg k^{\dw}$ (regime 2), see Eq. (1) of the main text.  

We start with $T \gg n^{\dw}$, so that all $S(k,T)$ are in the same regime $T \gg k^{\dw}$. Using Eq.~\eqref{eq:dist_prod},
\begin{align}
    \mathbb{P}(T_n \leq T) &\simeq \exp \left[-\text{cste.} \sum_{k=1}^{n-1} \frac{1}{k^{1-\epsilon_2}}\frac{1}{T^{y_2}} \right] \simeq \exp \left[-\text{cste.}\frac{n^{\epsilon_2}}{T^{y_2}} \right]
\end{align}
The scale-invariance of $T_n$ (Eq.~\eqref{eq:si_T}) implies that $\epsilon_2/y_2=\dw$. 

Next, for $T \ll n^{\dw} $,  $S(n,T)\propto n^{-1+\epsilon_1}T^{-y_1}$. By splitting the sum over $k$   between the regions $k^{\dw}  \ll T$ and $k^{\dw} \gg T$, we have
\begin{align}
    \mathbb{P}(T_n \leq T) \simeq & \exp \left[ -\text{cste.} \sum_{k=1}^{T^{1/\dw}} \frac{1}{k^{1-\epsilon_2}}\frac{1}{T^{y_2}} - \text{cste.} \sum_{k=T^{1/\dw}}^{n-1} \frac{1}{k^{1-{\epsilon_1}}}\frac{1}{T^{y_1}} \right] \nonumber\\
    \propto& \exp \left[-\text{cste.}  \frac{n^{\epsilon_1}}{T^{y_1}} \right] 
    =\exp \left[-\text{cste.}  \left(\frac{n^{\epsilon_1/y_1}}{T}\right)^{y_1} \right]. 
\end{align}
This result, together with Eq.~\eqref{eq:si_T}, yields $\epsilon_1/y_1=\dw=\epsilon_2/y_2$. In particular, the survival probability of the record-ages admits a scaling form in $T$ and $k$: 
\begin{align}
    S(k,T)=\frac{1}{k}\psi(T/k^{\dw}) \; . \label{eq:scale_inv}
\end{align}

 To proceed further, we look at the cumulative distribution of the increments ${T_{m+n}-T_m}$ in the limit $1 \ll n\ll m$ and $1 \ll T\ll m^{\dw}$, 
\begin{align}
    \mathbb{P}&(T_{m+n}-T_m \leq T) =\mathbb{P}\left(\sum_{k=m}^{n+m-1} \tau_k \leq T\right)\nonumber \\
    &\simeq \exp \left[-\text{cste.} \sum_{k=m}^{n+m-1} \frac{1}{k^{1-\epsilon_1}}\frac{1}{T^{y_1}} \right] \simeq  \exp \left[-\text{cste.}  \frac{n}{m^{1-\epsilon_1}T^{y_1}} \right]=\exp \left[-\text{cste.}  \left(\frac{n^{1/y_1}}{m^{(1-\epsilon_1)/y_1}T}\right)^{y_1} \right] \; .
\end{align}
Using finally Eq.~\eqref{eq:si_incT}, we obtain $y_1=1/\dw^0$ and $\frac{1-\epsilon_1}{y_1}=\dw^0-\dw$, leading to $\epsilon_1=\dw/\dw^0$. This provides the exponents $y_1$ and $\epsilon_1$ of regime 1 ($\tau_n \ll n^{\dw}$). We note that the algebraic decay of $S(n,\tau)$ \leo{holds} after the time $\tau_n$ such that
\begin{align}
    X_{\tau_n+T_{n}}-X_{T_{n}} &\propto T_{n}^{\alpha/2}\tau_n^{1/\dw^0} 
    \propto (n\Delta x)^{\dw \alpha/2}\tau_n^{1/\dw^0} 
    =(n\Delta x)^{1-\dw/\dw^0}\tau_n^{1/\dw^0}   \gg \Delta x \; .
\end{align}
This gives the lower bound $\tau_n\gg n^{\dw-\dw^0}$ in Eq.~(1) of the main text for regime 1.\\

For times $\tau_n \gg n^{\dw}$ (regime 2), the record age $\tau_n$ is much larger than the typical time needed to break all the previous records. At this time scale, the memory of the $n$ broken records no longer affects the algebraic time decay of $S(n, \tau )$, which is thus given by the usual persistence exponent $\theta = y_2$ (defined as $\mathbb{P}(T \geq \tau)\propto \tau^{-\theta}$ where $T$ is the time needed to reach a given value for the first time). Knowing that $\epsilon_2/y_2=\dw$, we obtain $\epsilon_2=\dw \theta$. 

\vspace{0.5cm}

By combining the results derived in this section, we finally obtain Eq.~(1) of the main text.

\subsection{General scaling criteria showing that the correlations between record ages are asymptotically irrelevant for their maximum}

\subsubsection{Effective independence criteria}

\leo{Here we show that in the calculation of the distribution of the maximum of the record ages, $M_n\equiv\max(\tau_0,\ldots,\tau_{n-1})$, the random variables $\tau_k$ can be treated as effectively independent. To do so, we extend the criteria of Ref.~\cite{carpentier2001}. Dividing the set $(\tau_0,\ldots,\tau_{n-1})$ in two subsets, $(\tau_0,\ldots,\tau_{n/2-1})$ and $(\tau_{n/2},\ldots,\tau_n)$, this criteria states that  the correlations between the  $\tau_k$
are  irrelevant if (a) the  typical  mean cross-correlation  between  the  subsystems is much smaller than (b)  the variance of the maximum of record ages}  of the whole system. 

Here, however, the random variables $\{\tau_k\}$ have diverging second moments, see Eq.~(1) of the main text. To overcome this difficulty, we extend the criteria of Ref.~\cite{carpentier2001} which uses variances of random variables and consider fractional moments of $\{\tau_k\}$ that converge, see below. Also, the original work of Ref.~\cite{carpentier2001} considers random variables that are Gaussian and identically distributed. Here we extend the approach to the distribution given in Eq.~(1) of the main text.

\subsubsection{General form of $\mathbb{P}(\tau_k \geq T)$}

We use dimensional analysis to obtain the scaling form of the tail distribution of record ages, without assuming their independence. There are no additional hypotheses on the RW processes beyond those  already requested: \\ (i) continuity, (ii) non-smoothness, and (iii) asymptotic scale-invariance.

We recall that the probability to reach level $(k+1)\Delta x=n+\Delta x$ starting from $k\Delta x=n$ at a time larger than $T$ is given by a function of microscopic cut-off $\Delta x$ (length), the level number $n$ (length) and $T$ (time). Based on dimensional analysis and scale-invariance of the process, the tail distribution of the record ages takes the functional form $\mathbb{P}(\tau_k \geq T)=F(\Delta x/n,T/n^{\dw})$. We are interested in the limit $\Delta x/n\ll 1$ and $T\sim n^{\dw}$, and consider (without restriction of generality) the limit behaviour 
\begin{align}
 \mathbb{P}(\tau_k \geq T)=F\left(\frac{\Delta x}{n},\frac{T}{n^{\dw}}\right)\sim \left(\frac{\Delta x}{n} \right)^\beta \psi(T/n^{\dw})  \label{eq:dist_tau_sc} 
\end{align}
with $\beta$ an exponent that we determine in the following.

\begin{figure}
    \centering
    \includegraphics[width=\columnwidth]{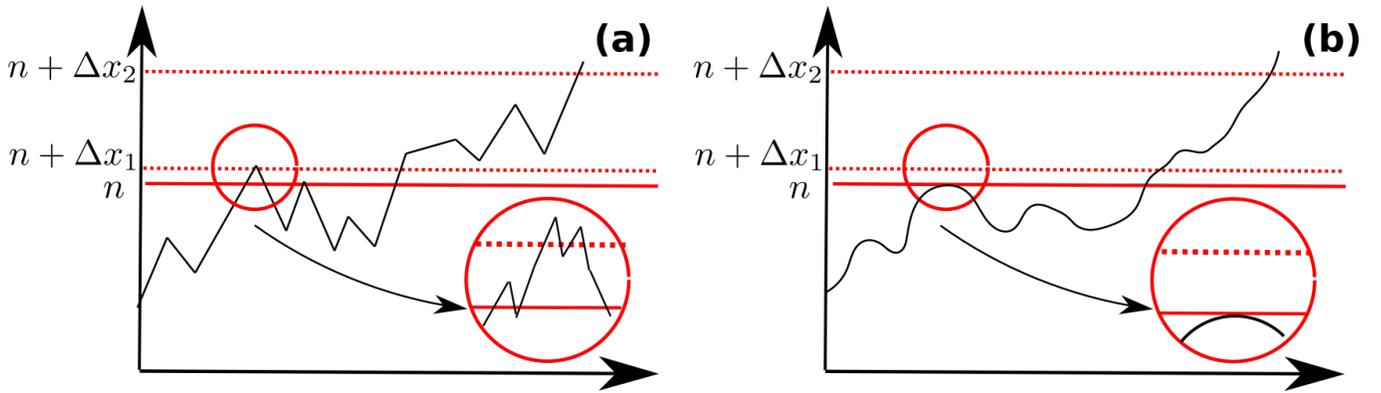}
    \caption{\leo{{\bf Non-smoothness versus smoothness.} We compare {\bf (a)} a non-smooth process to {\bf (b)} a smooth process by zooming on the trajectory that reaches level $n$ for the first time. For the non-smooth process the time needed to cross level $n+\Delta x_1$ (with a $\Delta x_1\ll n$ small enough) starting from level $n$ is significantly smaller than for the smooth process.}}
    \label{fig:smoothness}
\end{figure}

For continuous non-smooth processes, when reaching the $n$th level, the RW crosses the level $n$ infinitely often \cite{bray2013persistence}, see \ref{fig:smoothness}\textbf{a} for an illustration. In particular, this means that when $\Delta x$ is small compared to $n$, there is a time, much smaller than $T\sim n^{\dw}$, at which the trajectory crosses level $n+\Delta x$. Thus, the probability to reach level $n+\Delta x$ at a time larger than $T$ goes to $0$ when $\Delta x$ goes to $0$, which shows that the exponent $\beta$ in Eq.~\eqref{eq:dist_tau_sc} is strictly positive, $\beta>0$.

\subsubsection{\leo{Cross-correlations}}

We wish to find an upper bound of the typical correlations between different $\tau_k$ of the set $(\tau_0,\dots,\tau_{n})$,  for example, \oli{typically quantified} by the correlation between $\tau_{n/4}$ and $\tau_{3n/4}$. However, computing directly their covariance $\text{Cov}\left(\tau_{n/4},\tau_{3n/4} \right)$ would lead to a diverging result, because of a heavy-tailed $\tau_k$ distribution. Therefore, \oli{we will make use of}  the fractional powers of record ages, $\{\tau_k^q\}$, where the value  of $q$ allows converging second moments of these new variables.

An upper bound of the covariance $ \text{Cov}\left(\tau_{n/4}^q,\tau_{3n/4}^q \right)$ is obtained by using   the Cauchy-Schwarz inequality
\begin{align}
    \text{Cov}\left(\tau_{n/4}^q,\tau_{3n/4}^q \right) \leq \sqrt{\text{Var}(\tau_{n/4}^q)\text{Var}(\tau_{3n/4}^q)}.
\end{align}
The scaling behavior of each of the variances can be found based on Eq.~\eqref{eq:dist_tau_sc} for continuous non-smooth scale-invariant processes, leading to the moments ($k=1,2$)
\begin{align}
    \langle\tau_n^{kq}\rangle\propto n^{-\beta}\int_1^{\infty}\mathrm{d}\tau\,\tau^{qk-1}\psi(\tau/n^{\dw})\propto n^{-\beta+qk\dw}.
\end{align}
Note that for $\beta>0$, $\langle\tau_n^{2q}\rangle$ dominates $\langle\tau_n^{q}\rangle^2$, so that $\text{Var}(\tau_{n}^q)\lesssim\langle\tau_n^{2q}\rangle$. Using this, we get the following final estimation of the upper bound for the typical cross-correlations:
\begin{align}
    \text{Cov}\left(\tau_{n/4}^q,\tau_{3n/4}^q \right) \leq \sqrt{\text{Var}(\tau_{n/4}^q)\text{Var}(\tau_{3n/4}^q)}\propto n^{-\beta+2q\dw}.
\end{align}
This last inequality finally provides an upper bound of the typical correlations between the subsets $(\tau_0,\dots,\tau_{n/2-1})$ and $(\tau_{n/2},\ldots,\tau_n)$.

\subsubsection{\leo{Variance of max}}

The fluctuations of the maximum of the record ages  $M_n\equiv\max(\tau_0,\ldots,\tau_{n-1})$ are estimated by relying on  the variance of $M_n^q\equiv\max(\tau_0^q,\ldots,\tau_{n-1}^q)$. Based on Eqs.~\eqref{eq:si_T} and \eqref{eq:defTn} and on the self-consistently checked Eq.~\eqref{eq:summax},  the maximum's moments ($k=1,2$) can be written as: 
\begin{align}
     \langle M_n^{kq}\rangle\sim\int_1^{\infty}\mathrm{d}M\,M^{qk-1}(1-\Phi(M/n^{\dw}))\propto n^{qk\dw}.\label{eq:Maxtau}
 \end{align}
We choose $q$ such that the integral in Eq.~\eqref{eq:Maxtau} converges. Because $M_n$ is scale invariant ($M_n/n^{\dw}$ is a non-deterministic $n$ independent random variable), we finally have  
\begin{equation}
   \text{Var}\left(M_n^q \right)\propto\langle M_n^{2q}\rangle\propto n^{2q\dw}. 
\end{equation}

\subsubsection{\leo{Conclusion}}

We now can compare the typical cross-correlations to the fluctuations of the maximum,
\begin{align}
    \text{Cov}\left(\tau_{n/4}^q,\tau_{3n/4}^q \right) \leq \sqrt{\text{Var}(\tau_{n/4}^q)\text{Var}(\tau_{3n/4}^q)}\propto n^{-\beta+2q \dw} \ll n^{2q \dw} \propto\text{Var}\left(M_n^q \right). 
\end{align}
We conclude that the fluctuations of the maximum of record ages dominate the correlations between two record ages, so that the random variables $\tau_k$ can be considered as effectively independent.

\vspace{0.3cm}

We note that, a priori, the correlations between  $\{\tau_k\}$ are not negligible if one of the three hypotheses ((i) continuity, (ii) non-smoothness, and (iii) scale-invariance) breaks. While the absence of scale-invariance and of continuity would immediately invalidate calculations of this section, it is not evident for smooth processes. The necessity of the assumption of non smoothness originates from the following argument. When a realisation of  a smooth process just reaches the level $n$, in the following the trajectory can go back to position $x<n$ without crossing level $n$, see \ref{fig:smoothness}\textbf{b}. Thus, for smooth processes, the probability for a RW trajectory to reach level $n+\Delta x$ at a time $T$ for $\Delta x$  \oli{arbitrary small}  is finite, implying $\beta=0$ in Eq.~\eqref{eq:dist_tau_sc},  which invalidates the estimations after it.

\section{Non-Markovian random walks (RWs)}

\subsection{Definition of the non-Markovian RW models }

In this subsection, we present the non-Markovian random walk (RW) processes, which are used in Fig.~2 of the main text. These processes encompass the three classes of different statistical statistical mechanisms giving rise to a non-Markovian
evolution discussed in the main text and depicted in Fig.~1 of the main text. \ref{tab:recap} provides a summary of their characteristic parameters, namely $\dw$, $\alpha$, $\dw^0$ and $\theta$.

\begin{table*}[th!]
\begin{tabular}{@{\hspace{0.5cm}}c@{\hspace{0.5cm}}|@{\hspace{0.5cm}}c@{\hspace{0.5cm}}|@{\hspace{0.5cm}}c@{\hspace{0.5cm}}|@{\hspace{0.5cm}}c@{\hspace{0.5cm}}|@{\hspace{0.5cm}}c@{\hspace{0.5cm}}}
     Model & $\dw$ & $\alpha$ & $1/\dw^0$ & $\theta$  \\  \hline
     fBm & $1/H$ & $0$ & $H$ & $1-H$ \\
     qfBm  & $1/H$ & $0$ & $H$ & $\theta(H)$ (see \cite{krug1997persistence}) \\
     eRW & 2 & $0$ & 1/2 & $3/2-2\beta$ \\
     SATW & $2$ & $0$ & $1/2$ & $e^{- \beta}/2$ \\
     SESRW & $\frac{2+\kappa}{1+\kappa}$ & $0$ & $\frac{2+\kappa}{1+\kappa}$ &  $\approx 0.3$ \\
     TSAW  & $3/2$ & $0$ & $3/2$ & $1/3$ \\
     subALL & $3-a$ & $\frac{a-1}{3-a}$ & 1/2 & $(2-a)/(3-a)$\\
     supALL & $1+a$ & $\frac{1-a }{1+a}$ &$1/2$ & $a/(1+a)$ \\
     sBm & $2/\beta$ & $\beta-1$ & $1/2$ & $\beta/2$ 
\end{tabular}
\caption{ Summary of the non-Markovian models considered in this study and of their characteristic parameters.  \label{tab:recap} }
\end{table*}

\begin{enumerate}[label=(\textbf{\alph*})]
\item \textit{Fractional Brownian motion (fBm).} The fBm is a non-Markovian Gaussian process, with stationary increments. Thus, an fBm $X_t$ of Hurst index $H$ is defined by its covariance 
\begin{align}
    \text{Cov}\left(X_t,X_{t'} \right)= \frac{1}{2} \left(t^{2H} + t'^{2H}-|t-t'|^{2H} \right) \; .
\end{align}
The steps $\eta_t=X_t-X_{t-1}$ are called fractional Gaussian noise (fGn). 
Nowadays, the fBm is broadly spread and its implementations could be found in standard packages of python or Wolfram Mathematica. Besides, the survival probability of fBm is characterized by the persistence exponent $\theta=1-H$, which was derived in \cite{hansen1994,Ding1995,Maslov1994}.

\item\textit{Quenched fBm (qfBm).} This process is an extension of fBm to quenched initial conditions, which results in non-stationary increment statistics. In particular, it describes the height fluctuations under Gaussian noise of an initially flat interface. Then $X_t$ corresponds to the height of the interface at position $x=0$, $X_t=h(0,t)$, $h(x,t)$ following the Stochastic Differential Equation (SDE)
\begin{align}
    \partial_t h(x,t)=-\left(-\Delta \right)^{z/2} h(x,t) + \eta(x,t).
\end{align}
Here $\eta(x,t)$ is a Gaussian noise with possible spatial correlations. We solve numerically this SDE with a spatial discretization $\Delta x=1$ and a time discretization $\Delta t=0.1$. The system is initially flat, $h(x,t=0)=h_0$. The model at $z=2$ with space-independent noise is a non-stationary fBm of Hurst exponent $H=(1-1/z)/2=1/4$ which corresponds to the continuous limit of a solid-on-solid model \cite{Ryu1995}. The persistence exponent $\theta$ was numerically estimated  to be $\theta=1.55 \pm 0.02$ \cite{krug1997persistence}.

\item\textit{Elephant RW (eRW).}
This process is representative of interactions with its own trajectory. At time $t$, the step $\eta_t$ is drawn uniformly among all the previous steps $\eta_i$ ($i<t$) and is reversed with probability $\beta$. The persistence exponent was determined in \cite{barbier2020anomalous} to be $\theta=3/2-2\beta$.

\item\textit{Self-attractive walk (SATW).} This model is a
prototypical example of self-interacting RWs. In the SATW model \cite{sapozhnikov1994self,davis_1990,barbier2020anomalous,barbier2022self}, the RW at position $i$ jumps to a neighbouring site $j=i \pm 1$ with probability depending on the number of times $n_j$ it has visited site $j$, 
\begin{align}
    p(i\to j)=\frac{\exp \left[ -\beta H(n_j) \right]}{\exp \left[ -\beta H(n_{i-1}) \right]+\exp \left[ -\beta H(n_{i+1}) \right]},
\end{align}
where $H(0)=0$, $H(n>0)=1$ and $\beta>0$. It was shown in \cite{barbier2020anomalous} that the persistence exponent of this RW is given by $e^{-\beta}/2$.

\item[({\bf e-f})]\textit{Exponential self-repelling RW.} This is another example of self-interacting RW. In this model, the RW at position $i$ jumps to a neighbouring site $j=i \pm 1$ with probability depending on the number of times $n_j$ it has visited site $j$, 
\begin{align}
    p(i\to j)=\frac{\exp \left[ -\beta n_j^\kappa \right]}{\exp \left[ -\beta n_{i-1}^\kappa \right]+\exp \left[ -\beta n_{i+1}^\kappa \right]}
\end{align}
where $\kappa$ and $\beta$ are two positive real numbers. It was shown \cite{Ottinger_1985} that the walk dimension of such walks is given by $\dw=\frac{1+\kappa}{2+\kappa}$. In the case $\kappa=1$ (the True Self-Avoiding Walk, TSAW, \cite{Amit1983,Pietronero1983,Obukhov1983}), the persistence exponent is given by \cite{barbier2022self} $\theta=1/3$, while for $\kappa<1$ (the Sub-Exponential Self-repelling Walk, SESRW, \cite{Ottinger_1985,Toth1995}), its numerical estimation is $\theta \approx 0.3$.

\item[({\bf g-h})]\textit{Average L\'evy Lorentz gas (ALL).} This model is emblematic for RWs with spatially-dependent steps; Its different properties are described in \cite{Radice_2020,Radice2020_2}. We consider a RW on a $1d$ lattice with position dependent reflection or transmission probabilities $r(k)$ or $t(k)$. In the subdiffusive model (subALL), the transmission coefficient is taken to be 
\begin{align}
    t(k)= 
    \begin{cases}
       \frac{a \sin \left( \pi a \right) \zeta(1+a)}{2\pi |k|^{1-a}}& \mbox{ if } |k|>0 \\
    1/2 &\mbox{ otherwise } 
    \end{cases}
\end{align}
In the continuous setting, this is equivalent to a space dependent diffusion coefficient for $0<a<1$, $D_a(x)=\left(4\Lambda |x|^{1-a}-2 \right)^{-1}$ where $\Lambda=\frac{\pi}{a\sin(\pi a)\zeta(1+a)}$. The persistent exponent is given by  $\theta=1-1/(3+a)$ \cite{Radice2020_2}. In the superdiffusive model (supALL), the reflection coefficient is taken to be 
\begin{align}
    r(k)=
    \begin{cases}
    \frac{a \sin \left( \pi a \right) \zeta(1+a)}{2\pi |k|^{1-a}} & \mbox{ if } |k|>0 \\
    1/2 &\mbox{ otherwise.}
    \end{cases}
\end{align}
In the continuous setting, this is equivalent to a space dependent diffusion coefficient for $0<a<1$, $D_a(x)=\Lambda |x|^{1-a}-1/2 $ where $\Lambda=\frac{\pi}{a\sin(\pi a)\zeta(1+a)}$. The persistent exponent is given by $\theta=1-1/(1+a)$ \cite{Radice2020_2}. \\

\item[({\bf i})]\textit{Scaled Brownian motion (sBm).}
The sBm model represents RWs with time-dependent steps \cite{Safdari2015,Lim2002, jeon2014scaled,He2008}. Starting from $X_t^0$ a process with i.i.d. symmetric jumps $\eta_i^0$ with finite variance, we define the scale process of parameter $\beta$ by $X_t\equiv X_{\left\lfloor t^\beta \right\rfloor }^0$, or equivalently $\eta_t = \sum\limits_{k=\left\lfloor (t-1)^\beta \right\rfloor}^{\left\lfloor t^\beta \right\rfloor-1}\eta_k^0$. A second way to define the sBm on discrete times is to consider steps $\eta_t$ following a binomial distribution of parameters $(N_t,1/2)$ where $N_t$ is Poisson distributed of average $\lambda(t) =t^{\beta-1}$ (the one used in \ref{fig:Averages}). In the continuous setting, it amounts to a time-dependent overdamped Langevin equation, 
\begin{align}
    \frac{dX_t}{dt}=\sqrt{2 D \beta t^{\beta-1}}\eta_t,
\end{align}
where $\eta_t$ is a white noise of unit variance. The persistence exponent of sBm is  $\theta=\beta/2$ \cite{Lim2002}. 

The sBm model allows one to compute the distribution of $\tau_{x_0} \equiv T_{x_0+\Delta x}-T_{x_0}$ analytically in the continuous limit,

\begin{align}
  \mathbb{P}(\tau_{x_0} \geq \tau)&=  \int_0^\infty dT_0  \mathbb{P}(T^0_{x_0}=T_0) \mathbb{P}(T^0_{\Delta x} \geq (\tau+T_0^{1/\beta})^\beta-T_0) \nonumber \\
  &= \int_0^\infty dT_0  \frac{x_0}{\sqrt{2 \pi T_0^3}}\exp\left[-\frac{x_0^2}{2T_0} \right] \text{erf}\left(  \frac{\Delta x \sqrt{2}}{\sqrt{(\tau+T_0^{1/\beta})^\beta-T_0}} \right) \label{eq:tail_scaledBM}
\end{align}
We plot this exact formula in Fig. 2 of the main text ($x_0=n$, $\Delta x=1$) and in 
 \ref{fig:ScaleInv} ($x_0=m$, $\Delta x=n$). In the case of small $\Delta x$, 
\begin{align}
  \mathbb{P}(\tau_{x_0} \geq \tau) &\sim\frac{2 \Delta x}{\pi } \int_0^\infty dT_0  \frac{x_0}{ \sqrt{ T_0^3}}\exp\left[-\frac{x_0^2}{2T_0} \right] \frac{1}{\sqrt{(\tau+T_0^{1/\beta})^\beta-T_0}} \propto 
  \begin{cases}
 \frac{\Delta x}{\sqrt{\tau}x_0^{1-1/\beta}} &\text{ for } \tau \ll x_0^{2/\beta}, \\
\frac{ \Delta x}{ \tau^{\beta/2}} &\text{ for } \tau \gg x_0^{2/\beta},
  \end{cases}
\end{align}
in full agreement with the central result of this work, Eq. (1) of the main text.
  \end{enumerate}

\subsection{Systematic numerical check of the  scale-invariance of the time increments}

\begin{figure}[th!]
    \centering
    \includegraphics[width=\columnwidth]{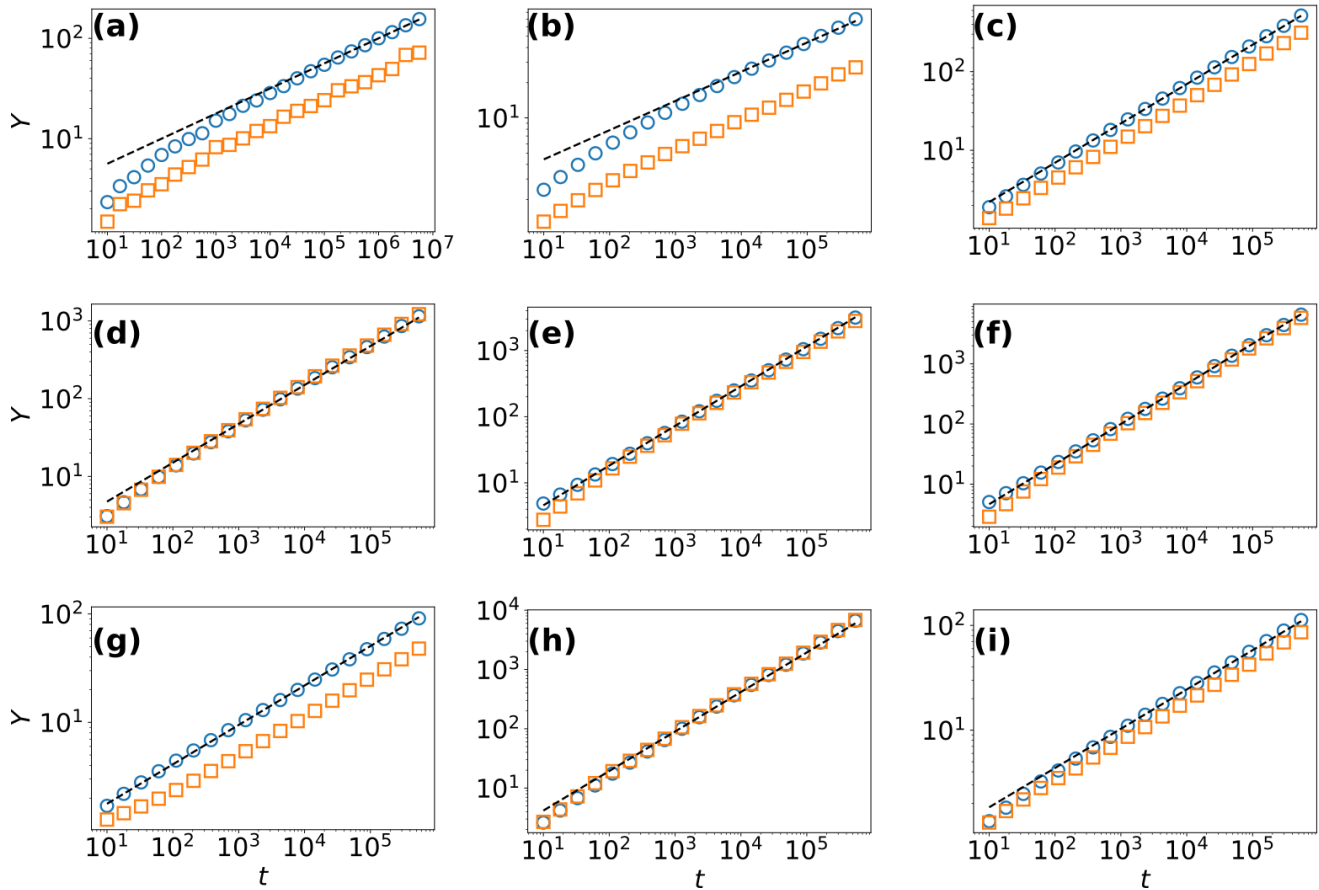}
    \caption{{\bf Average and standard deviation of number of records.} Each subpanel represents the average \leo{(blue circles)} and standard deviation \leo{(orange squares)} of the number of records reached at time $t$ compared to the scaling expectation $\propto t^{1/\dw}$ \leo{(dashed lines)} for \textbf{(a)} fractional Brownian motion (fBm) of Hurst exponent $H=0.25=1/\dw$, \textbf{(b)} quenched fBm (qfBm) of Hurst exponent $H=0.25=1/\dw$,  \textbf{(c)} elephant RW (eRW) of parameter $\beta=0.25$ such that $\dw=2$, \textbf{(d)} Self-Attractive Walk (SATW) of parameter $\beta=1$, such that $\dw=2$, \textbf{(e)} Sub-Exponential Self-Repelling Walk (SESRW) of parameter $\beta=1$ and $\kappa=0.5$ such that $\dw=5/3$,  \textbf{(f)} True Self-Avoiding Walk (TSAW) of parameter $\beta=1$ such that $\dw=3/2$,  \textbf{(g)} Subdiffusive Average L\'evy Lorentz (subALL) of parameter $a=0.25$ such that $\dw=2.75$, $\dw^0=2$, \textbf{(h)} Superdiffusive Average L\'evy Lorentz (supALL) of parameter $a=0.5$ such that $\dw=3/2$, $\dw^0=2$, and \textbf{(i)} Scaled Brownian motion (sBm) of parameter $\beta=0.75$ such that $\dw=8/3$.}
    \label{fig:Averages}
\end{figure}

\begin{figure}[th!]
    \centering
    \includegraphics[width=\columnwidth]{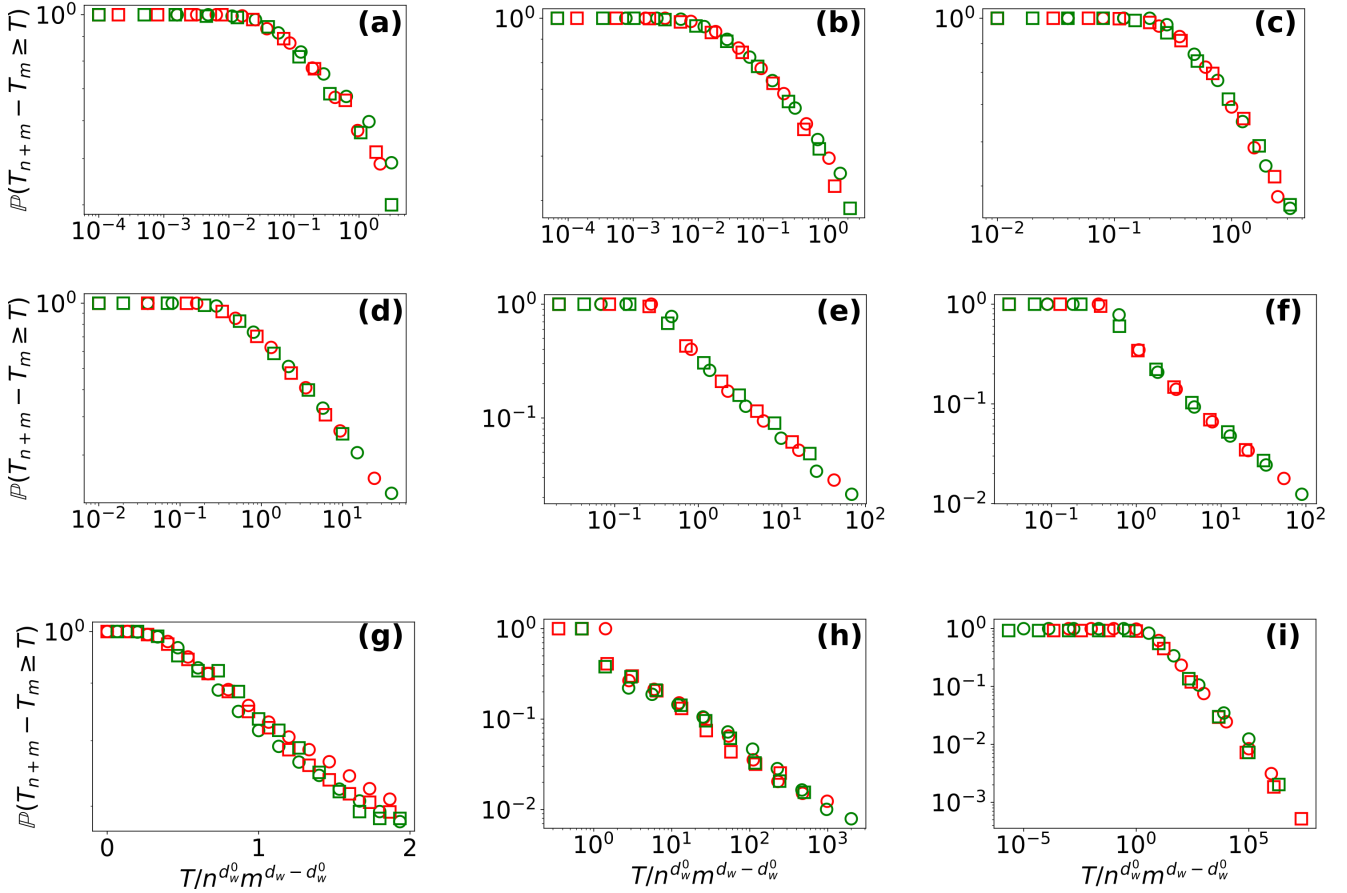}
    \caption{{\bf Scaled distribution of the time increments $T_{m+n}-T_m$.} Each subpanel represents the simulated tail distribution of the random variable $T_{m+n}-T_m$ as a function of $T/n^{\dw^0}m^{\dw-\dw^0}$ for different values of $n$ and $m$ for \textbf{(a)} fBm of Hurst exponent $H=0.25=1/\dw$ ($n=5$ and $10$, $m=50$ and $100$) \textbf{(b)} qfBm of Hurst exponent $H=0.25=1/\dw$  ($n=5$ and $10$, $m=50$ and $100$) \textbf{(c)} eRW of parameter $\beta=0.25$ such that $\dw=2$ ($n=5$ and $10$, $m=50$ and $100$) \textbf{(d)} SATW of parameter $\beta=1$, such that $\dw=2$ ($n=5$ and $10$, $m=100$ and $500$) \textbf{(e)} SESRW of parameter $\beta=1$ and $\kappa=0.5$ such that $\dw=5/3$ ($n=5$ and $10$, $m=100$ and $500$) \textbf{(f)} TSAW of parameter $\beta=1$ such that $\dw=3/2$ ($n=5$ and $10$, $m=100$ and $500$) \textbf{(g)} subALL of parameter $a=0.25$ such that $\dw=2.75$, $\dw^0=2$ ($n=10$ and $20$, $m=400$ and $800$) \textbf{(h)} supALL of parameter $a=0.5$ such that $\dw=3/2$, $\dw^0=2$ ($n=5$ and $10$, $m=50$ and $100$) \textbf{(i)} Exact tail distribution for sBm of parameter $\beta=0.75$ such that $\dw=8/3$ ($n=10$ and $100$, $m=1000$ and $10000$). Increasing values of $m$ are represented respectively by red and green symbols, while increasing values of $n$ are represented respectively by circles and squares. Times $T_{m}>10^6$ are discarded to have finite computation times.}
    \label{fig:ScaleInv}
\end{figure}

Based on the RW models described above, we confirm the scale invariance of the following random variables : 
\begin{itemize}
    \item {\it The number of records.} We show in  \ref{fig:Averages} that the number of records at time $t$ is indeed scale invariant at large times as its average and standard deviation grow as expected as $t^{1/\dw}$ for all non-Markovian process considered. 
     \item {\it The time-increments between records.} \ref{fig:ScaleInv} provides the distributions of the time increments $T_{m+n}-T_m$ and checks their scale invariance with respect to $m$ and $n$; namely, that $(T_{m+n}-T_m)/n^{\dw^0}m^{\dw-\dw^0}$ is independent of $m$ and $n$ for $n \ll m$.
\end{itemize}

\clearpage

\subsection{Systematic numerical check of the asymptotic independence of record ages} 

\begin{figure}[th!]
    \centering
    \includegraphics[width=\columnwidth]{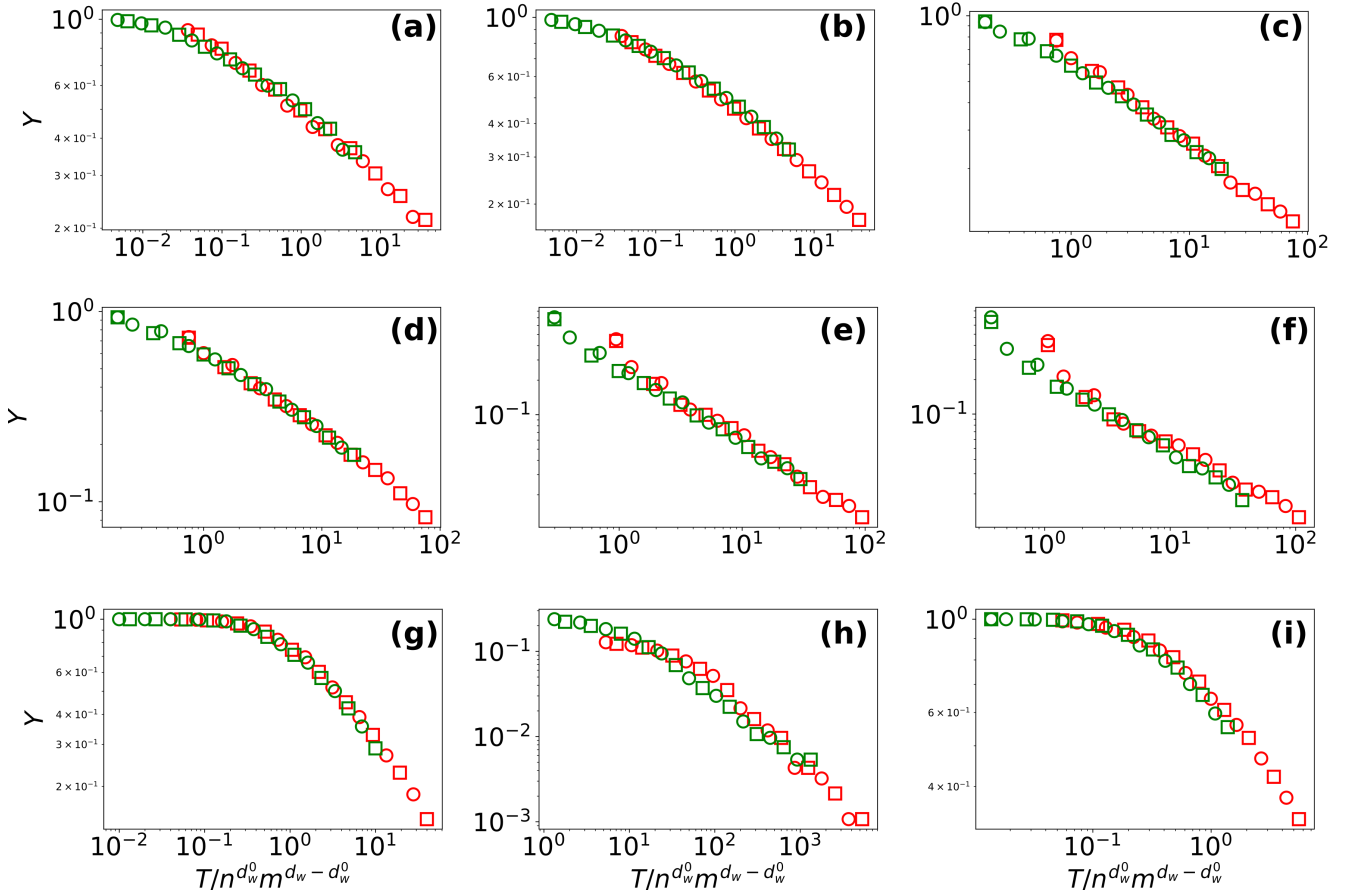}
    \caption{{\bf Scaled distribution of the record maximum with and without the independence approximation.} Each subpanel represents the tail distribution of the random variable $\max(\tau_m,\ldots, \tau_{n+m-1})$, $\mathbb{P}(\tau_k> T,~ k=m,\ldots, n+m-1)=1-\mathbb{P}(\tau_k\leq T,~ k=m,\ldots, n+m-1)$ (circles), and the product tail distribution of $\tau_m,\ldots,\tau_{n+m-1}$, $1-\prod_{k=m}^{m+n-1}\mathbb{P}(\tau_k\leq T)$ (squares), as a function of $T/n^{\dw^0}m^{\dw-\dw^0}$ with $n=2$ (red), $n=4$ (green) and $m=50$ for \textbf{(a)} fBm of Hurst exponent $H=0.25=1/\dw$,  \textbf{(b)} qfBm of Hurst exponent $H=0.25=1/\dw$  ($n=2$ and $4$), \textbf{(c)} eRW of parameter $\beta=0.25$ such that $\dw=2$, \textbf{(d)} SATW of parameter $\beta=1$, such that $\dw=2$, \textbf{(e)} SESRW of parameter $\beta=1$ and $\kappa=0.5$ such that $\dw=5/3$, \textbf{(f)} TSAW of parameter $\beta=1$ such that $\dw=3/2$, \textbf{(g)} SubALL of parameter $a=0.25$ such that $\dw=2.75$, $\dw^0=2$, \textbf{(h)} SupALL of parameter $a=0.5$ such that $\dw=3/2$, $\dw^0=2$, and \textbf{(i)} sBm of parameter $\beta=0.75$ such that $\dw=8/3$. Times $T_{m+n}>10^6$ are discarded to have finite computation times.}
    \label{fig:IndepApprox}
\end{figure}

\begin{figure}[th!]
    \centering
    \includegraphics[width=\columnwidth]{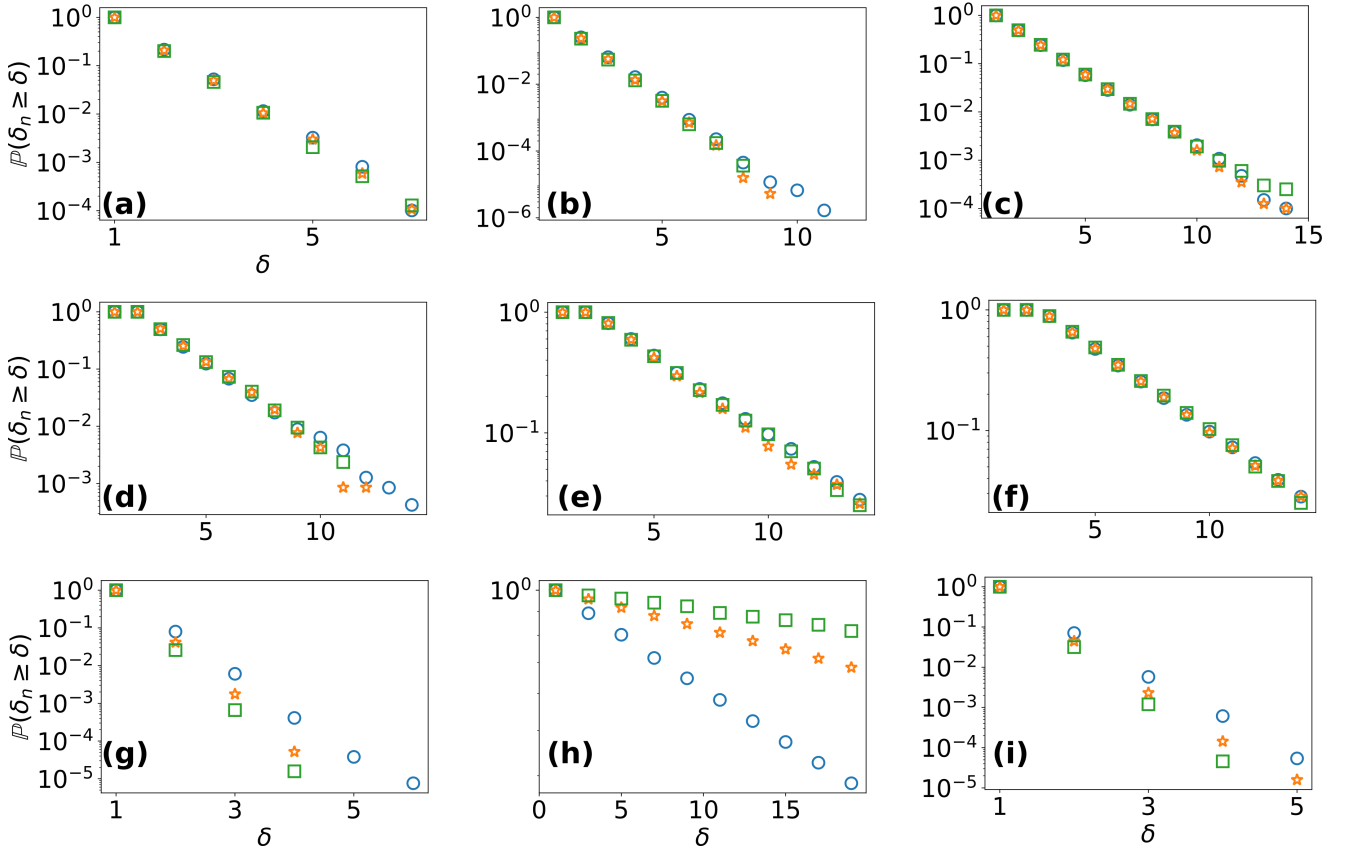}
    \caption{{\bf Distributions of record runs.} Each subpanel represents the distribution of a record run $\delta_n$ for different values of the number $n$ of previous record runs for \textbf{(a)} fBm of Hurst exponent $H=0.25=1/\dw$, \textbf{(b)} qfBm of Hurst exponent $H=0.25=1/\dw$,  \textbf{(c)} eRW of parameter $\beta=0.25$ such that $\dw=2$, \textbf{(d)} SATW of parameter $\beta=1$, such that $\dw=2$, \textbf{(e)} SESRW of parameter $\beta=1$ and $\kappa=0.5$ such that $\dw=5/3$, \textbf{(f)} TSAW of parameter $\beta=1$ such that $\dw=3/2$, \textbf{(g)} subALL of parameter $a=0.25$ such that $\dw=2.75$, $\dw^0=2$, \textbf{(h)} supALL of parameter $a=0.5$ such that $\dw=3/2$, $\dw^0=2$, and \textbf{(i)} sBm of parameter $\beta=0.75$ such that $\dw=8/3$. All distributions are computed for $n=10$, $25$ and $50$ record runs (blue circles, orange stars and green squares). Times to reach $n$ record runs larger than $10^6$ are discarded to have finite computation times.}
    \label{fig:Records}
\end{figure}

Based on the RW models described above, we make systematic tests of the independence hypothesis between record ages: 
\begin{itemize}
     \item In \ref{fig:IndepApprox}, we check the effective independence hypothesis used in Eq.~(3) of the main text, by comparing $\mathbb{P}(\max(\tau_m,\ldots, \tau_{m+n-1}) \leq T)$ with $\prod_{k=m}^{n+m-1} (1-S(k,T))$ for various values of $n$ in the regime $n^{\dw}$ and $T$ small in comparison to $m^{\dw}$. The functional forms being the same for both distributions even for different values of $n$ (by rescaling with $n$ and $m$), this confirms numerically the approximation for all the non-Markovian models considered.
    \item In \ref{fig:Records}, we display the probability to make at least $\delta$ successive records (event called a record run of length $\delta$) when $n$ record runs have been performed. In other words, we look at the joint distribution $\lbrace \tau_n=1,\ldots,\tau_{n+k}=1 \rbrace$, which shows an exponential decay in the correlations between successive record ages, and thus provides an additional numerical check of the independence of $\tau_k$. We note that the time decay rate of the exponential varies with $n$ for aging RWs,  as expected from the dependency on the number $n$ of records  of the early time regime for the record age $\tau_n$.
\end{itemize}

\clearpage

\section{Data analysis}
We provide a comprehensive description of the datasets used in this study, along with the general methodology that yields the results presented in the main text. Furthermore, we include complementary datasets that provide additional confirmation of our findings, including cases involving aging time series.

\subsection{Details on the datasets used in the main text}
Here, we present the datasets used in the main text:
\begin{enumerate}[label=(\textbf{\alph*})]
    \item \textit{Elbe river discharge ($m^3/s$).} We consider the daily mean debit of the Elbe river measured in Dresden \cite{GRDC_dresden}. It was observed that this quantity presents correlation which can be modeled by subdiffusive non-Markovian RWs \cite{zhang2008}. In this time series, we obtain $H\approx 0.14$ by application of the DMA method \cite{holl2019,alessio2002second}, see below.
    \item \textit{Volcanic soil temperature ($^o C$).} It was shown in \cite{Crescenzo2022,sabbarese2020} that the soil temperature monitored in the volcanic caldera of the Campi Flegrei area in Naples follows an fBm of parameter $H \approx 0.42$ once the data have been detrended by removing the linear trends between two temperature extrema (between two solstices). Here we also detrend the data by removing the same linear seasonal trends. We display the data measured at the Monte Olibano (OLB) site. 
    \item \textit{Trajectories of microspheres in agarose gel $(nm)$.} 
    The trajectories \cite{Krapf2019} represent the $2d$ motion of 50-nm polystyrene microspheres in agarose
    hydrogel (we consider that $x$ and $y$ displacement are i.i.d., giving us $2$ independent $1d$ trajectories). There are $20$ trajectories of $2000$ frames which were analyzed in \cite{Krapf2019} who obtained $1/\dw \approx 0.43$.
    \item \textit{Motion of amoeba intracellular vacuoles $(pixels=106nm)$.} We consider vacuole intracellular trajectories inside the amoeba in a $2d$ plane (we consider that $x$ and $y$ displacement are i.i.d., giving us $2$ independent $1d$ trajectories) of at least $2048$ frames. It was estimated in \cite{Krapf2019} that the walk dimension verifies $1/\dw \approx 0.67$.
     \item \textit{Trajectories of telomeres ($\mu m$).} We use $2d$ trajectories \cite{Krapf2019} of telomeres in the nucleus of untreated
    U2OS cells obtained in Ref. \cite{stadler2017} (we consider that $x$ and $y$ displacement are i.i.d., giving us $2$ independent $1d$ trajectories). Similarly to Ref. \cite{Krapf2019}, we only consider trajectories where the mean-square displacement grows as $t^{0.5 \pm 0.05}$, which corresponds to $1/\dw \approx 0.25$.
    \item \textit{DNA RW on the Homo sapiens $\beta$-myosin heavy chain (HUMBMYH7).} It was observed in \cite{Peng1994,peng1992long} that the process is a RW with long-range correlations of Hurst exponent $H \approx 0.67$.  We estimate and remove the bias $\hat{v}=\frac{1}{N}\sum_{t=1}^N \eta_t$ in the data by replacing $\eta_t$ by $\eta_t-\hat{v}$.
     \item \textit{Cumulative London air temperature  $(^o C.$day$)$.} In this case, the temperature fluctuations (as for \textbf{(b)}, we remove the linear trends between two solstices) are fractional Gaussian noise (fGn), in agreement to what was observed in \cite{brody2002}, of Hurst exponent $H\approx 0.8$ obtained via the DMA method \cite{holl2019,alessio2002second}, see below. We consider the cumulative temperature fluctuation, which is then fBm. This quantity is of a particular interest in the studies of derivative pricing (see \cite{brody2002}).
     \item \textit{Cumulative Ethernet traffic $(10~ bytes.ms)$.} The dataset represents the number of packets going through an Ethernet cable every 10 $ms$ at the Bellcore Morristown Research and Engineering facility \cite{Fowler1991}. In particular, in \cite{leland1993}, it was shown that the process is a fGn of dimension $H\approx 0.8$. As for \textbf{(e)}, the cumulative number of requests up to a given time $t$ gives a non-Markovian process. We detrend the cumulative data as for \textbf{(g)} by removing the estimated bias $\hat{v}$ at every step. We display the measurement performed in August 1989.
\end{enumerate}
\subsection{Characterization and parametrization of the data used in the main text}
In this section we provide the method developed to determine the walk dimension of the time series presented in the main text as well as numerical checks of their stationarity.\\

In order to obtain the walk dimension $\dw$ in a time series, one applies the celebrated Detrending Moving Average (DMA) \cite{alessio2002second,holl2019} method, which consists in evaluating the typical fluctuations in a window of size $\ell$ regardless of any bias or deterministic trend. More precisely, for a dataset $(X_t)_{t=0,\ldots,N}$, we consider the windows of size up to $\ell_\text{max}$, compute the window averages $x_t^\ell=\frac{1}{\ell}\sum_{i=0}^{\ell-1}X_{t-i}$, and the typical fluctuation for a window of size $\ell$, $F(\ell)=\sqrt{\frac{1}{N-\ell_\text{max}}\sum_{t=\ell_\text{max}}^N (X_t-x_t^\ell)^2}$. When several trajectories are available, we consider the average fluctuation over all the trajectories (for telomeres, vacuoles and microspheres in agarose data). If the data behave as a RW of walk dimension $\dw$, then $F(\ell) \propto \ell^{1/\dw}$.
We obtain the value of $1/\dw$ via the DMA method (first two lines in \ref{fig:DMAMSD}). Then, we compare the exponent with that obtained from the Mean Square Displacement (MSD). As one can see in the last two lines of \ref{fig:DMAMSD}, the MSD has the  algebraic growth predicted by the DMA method, which indicates that the deterministic trends have been removed correctly. 

In order to check that the data are stationary, we compare the MSD obtained from the increments $\lbrace x_t=X_{t+T}-X_T \rbrace_{T\leq N/4,t}$ in the first quarter of the data and the increments $\lbrace x_t=X_{t+T}-X_T \rbrace_{3N/4 \leq T,t}$ in the last quarter of the data.  Indeed, for all datasets the MSD in the two sub-intervals of the data have similar growth, i.e. the aging exponent is $\alpha=0$. We note that for the river flow dataset there is difference a constant prefactor. This transient aging explains the small deviations in Fig.~3 {\bf (a')} of the main text from the behavior of a stationary process characterized by the persistence exponent $\theta=1-1/\dw$. However, since the walk dimension is not changed, $\alpha=0$ and the record age exponent is still $1/\dw$ for this dataset.

Record ages are obtained by starting the subtrajectories at values of $t$ equally spaced at intervals at least $200$ time steps long, and observing successive records occurring in the subtrajectory. First return times are obtained by starting the subtrajectories at any value of time.

\begin{figure*}[th!]
    \centering
    \includegraphics[width=\columnwidth]{DMAMSD.pdf}
    \caption{ {\bf Characterization of the data used in Fig. 3 of the main text:}\\
    {\bf (a)} river discharge,  {\bf (b)} volcanic soil temperature, {\bf (c)} motion of microspheres in a gel, {\bf (d)} motion of vacuoles inside an amoeba, {\bf (e)} motion of telomeres, {\bf (f)} DNA RW, {\bf (g)} cumulative air temperature, and   {\bf (h)} Ethernet cumulative requests.\\
    Top subfigures  {\bf (a)}- {\bf (h)} show $F(\ell)$ obtained via the DMA method. The linear fit on the log-log plot is represented by a black dashed line.\\
    Bottom subfugures {\bf (a$^\prime$)}- {\bf (h$^\prime$)} show the MSD $\left\langle x_t^2 \right\rangle $ computed from the first (blue circles) and last (orange squares) quarters of the data. Black dashed line stands for the algebraic growth $t^{2/\dw}$ where $\dw$ was obtained via the DMA method.}
    \label{fig:DMAMSD}
\end{figure*}

\subsection{Analysis of complementary datasets}

We additionally conducted an analysis \mx{of} the following complementary datasets to further demonstrate the broad applicability of our results:

\begin{enumerate}[label=(\textbf{\alph*})]
    \item \textit{Volcanic soil temperature $(^oC)$.} Another dataset (measured at the Monte Sant’Angelo site) of soil temperatures monitored in the volcanic caldera of the Campi Flegrei area in Naples. It was shown in \cite{sabbarese2020,Crescenzo2022} that \mx{it} is an fBm of parameter $H \approx 0.4$ once the data have been detrended by removing the linear trends between two temperature extrema (between two solstices). Here we also detrend the data by removing the same linear seasonal trends.
    \item \textit{Cumulative air temperature at Mont\'elimar $(^oC.$day$)$.} As for the soil temperature in \cite{Crescenzo2022}, we remove the linear trends between two solstices. In this case, the temperature fluctuations are fractional Gaussian noise (fGn) and not fBm, in agreement to what was observed in \cite{brody2002} for the data at London.  We obtain a Hurst exponent $H\approx 0.8$ via the DMA method.
    \item \textit{Rh\^one river discharge $(m^3/s)$.} We consider the daily mean debit of the Rh\^one measured at the Sault Brenaz station \cite{GRDC}. It was observed that this quantity presents correlation\mx{s} which can be modeled by subdiffusive non-Markovian RW \cite{zhang2008}. In this time series, we obtain $H\approx 0.21$ by applications of the DMA method.
    \item \textit{DNA RW.} We consider the human T-cell receptor $\alpha$/$\delta$ sequence from the GenBank data base (HUMTCRADCV). It was observed in \cite{Peng1994,peng1992long} that the process is a RW with long-range correlations of Hurst exponent $H \approx 0.61$.  We estimate and remove the bias $\hat{v}=\frac{1}{N}\sum_{t=1}^N \eta_t$ in the data by replacing $\eta_t$ by $\eta_t-\hat{v}$.
    \item \textit{Cumulative Ethernet traffic $(10~bytes.ms)$.} The dataset represents the number of packets going through an Ethernet cable every 10 $ms$ at the Bellcore Morristown Research and Engineering facility \cite{Fowler1991}. In particular, in \cite{leland1993}, it was shown that the process is a fGn of dimension $H\approx 0.84$. This is why we consider the cumulative number of requests up to a given time $t$. We detrend the cumulative data in the same manner as the DNA RW by removing the estimated bias $\hat{v}$ at every step. We display the data measured in October 1989.
\end{enumerate}
All considered complementary datasets support our theoretical results.

\begin{figure*}[th!]
    \centering
    \includegraphics[width=\columnwidth]{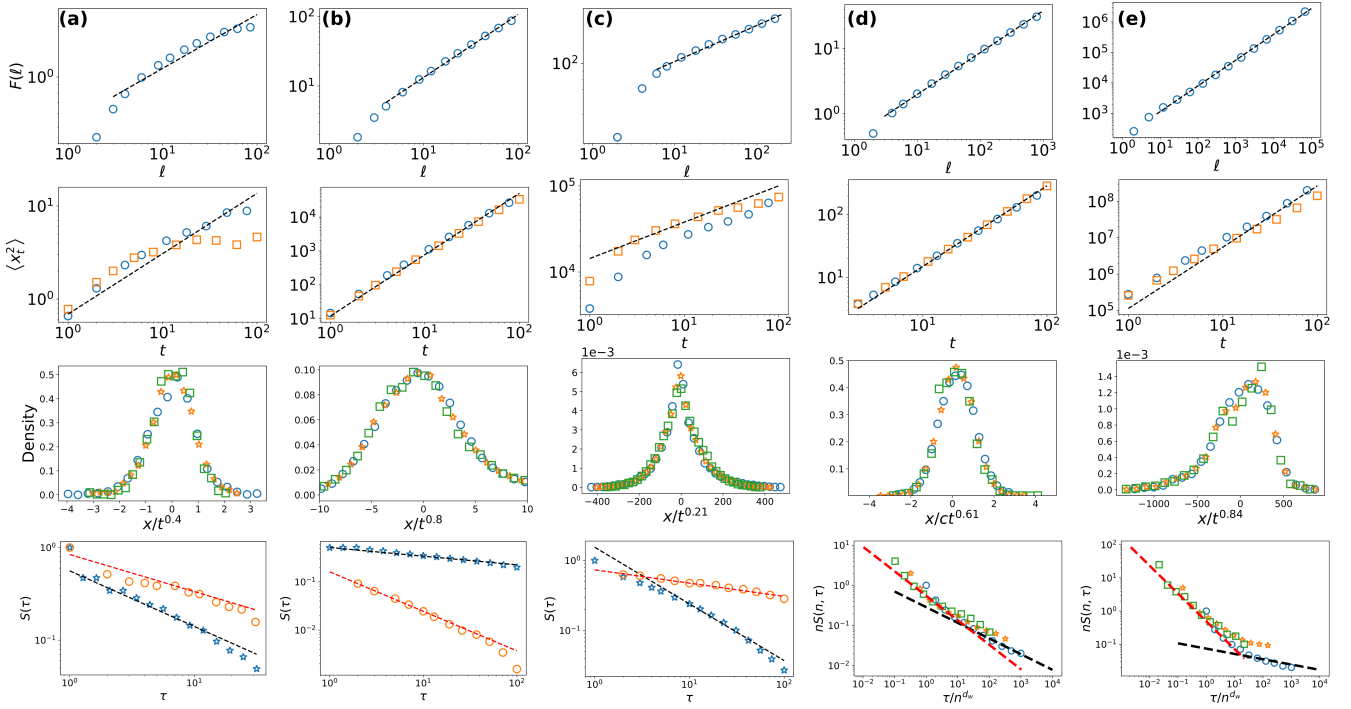}
 \caption{ {\bf Analysis of record ages for non-Markovian RWs: theoretical predictions (lines) vs real time observations (symbols):}  
    {\bf (a)} volcanic soil temperature (Monte Sant’Angelo site), {\bf (b)} cumulative air temperature (Mont\'elimar), {\bf (c)} river discharge (Rh\^one), {\bf (d)} DNA RW (HUMTCRADCV), and {\bf (e)} Ethernet cumulative requests (October 1989).\\
    First line,  $F(\ell)$ obtained via the DMA method and linear fit to the log-log representation shown by black dashed line.\\
    Second line, check of the MSD $\left\langle x_t^2 \right\rangle $ computed from the first (blue circles) and last (orange squares) quarters of the data. Black dashed line shows linear fit to the log-log representation of the MSD data, given the same exponent $2/\dw$. \\
    Third line, distribution of the increment $x_t=X_{t+T}-X_{T}$ at different times $t$ normalised by $t^{1/\dw}$, where $\dw$ was obtained via the DMA method for: {\bf (a)} $t=5$, $10$ and $20$ {\bf (b)} $t=5$, $10$ and $20$ {\bf (c)} $t=10$, $20$ and $40$ {\bf (d)} $t=20$, $40$ and $80$ {\bf (e)} $t=500$, $1000$ and $2000$. Increasing values of times are represented successively by blue circles, orange stars and green squares.\\
    Fourth line, {\bf (a)}-{\bf (c)} statistics of the time to first reach the initial value in the sub interval (blue stars) and the statistics of the records (regardless of the number $n$ of records, orange circles) and {\bf (d)}-{\bf (e)} rescaled tail distribution of record ages $\tau_n$  for different values of the number of records $n$ ($n=1,$ $2$ and $4$ for {\bf (d)} and $n=1$, $5$, $25$ for {\bf (e)}). The black dashed line represents the algebraic decay $\tau^{-\theta}$ while the red dashed line stands for the algebraic decay $\tau^{-1/\dw}$. 
    }   
    \label{fig:Data2}
\end{figure*}

\clearpage

\subsection{Datasets displaying aging of the increments}

We also analysed the following complementary datasets which present aging in the increments:

\begin{enumerate}[label=(\textbf{\alph*})]
    \item \textit{Single cell displacement on a $1d$ medium (half-pixels=$0.65\mu m$).} We analyze the motion of MDCK (Madin-Darby Canine Kidney) epithelial cells on micro-contact-printed $1d$ linear tracks of fibronectin obtained in \cite{Alessandro2021}, who found that the cells perform a Persistent Self-Attractive Walk (PSATW, which is a generalization of the SATW with a finite correlation length) motility behaviour, such that $\dw=\dw^0=2$. Because of the (transient, $\alpha=0$) aging in the data, the persistence exponent $\theta$ is different from $1/2$.  
    \item \textit{Single cell displacement on a $2d$ medium (pixels=$1.3\mu m$).} We analyze the motion of MDCK epithelial cells on a $2d$ substrate obtained in \cite{Alessandro2021}, who found that the cells show a PSATW  motility behaviour, such that $\dw=3$ and $\dw^0=2$ in agreement with theoretical and numerical observations for this type of model in $2d$ \cite{Ordemann2001,Foster_2009}. We consider that motion in $x$ and $y$ are i.i.d. and thus form two independent $1d$ trajectories. Because of the aging in the data, the persistence exponent $\theta$ is different from $1/2$.
\end{enumerate}

\begin{figure*}[th!]
    \centering
    \includegraphics[width=0.5\columnwidth]{SAT.pdf}

\caption{ {\bf Analysis of record ages for non-Markovian RWs: theoretical predictions (lines) vs  experimental data displaying aging of the increments (symbols).}
Cell motility on a {\bf (a)} $1d$ and {\bf (b)} $2d$ micropatterned surfaces.\\
First line, MSD $\left\langle x_t^2 \right\rangle $ computed at early times (blue circles, to compare with the red line $\propto t^{2/\dw}$) and at latter times (orange squares, to compare with the black line $\propto t^{2/\dw^0}$).\\ 
Second line, distribution of the increment $x_t=X_{t+T}-X_{T}$ at different times $t$ normalised by $t^{1/\dw}$, where $\dw$ was derived in \cite{Alessandro2021}, for $t=5$, $10$ and $20$ ($T<50$). Increasing values of times are represented successively by blue circles, orange stars and green squares.\\
Third line,  statistics of the time to first reach the initial value in the sub interval (blue stars) and the statistics of
the records (regardless of the number $n$ of records, orange circles). The black dashed line stands for the algebraic decay $\tau^{-\theta}$, where $\theta$ is estimated to be {\bf (a)} $\theta\approx 0.7$ and {\bf (b)} $\theta\approx 0.4$, and the red line represents the algebraic decay $\tau^{-1/\dw^0}$.}
    \label{fig:SAT}
\end{figure*}
\ref{fig:SAT} shows the analysis of datasets {\bf (a)} and {\bf (b)}. Note that, because of aging, the persistence exponent $\theta \neq 1-1/\dw$. We conclude that these complex examples also support our theory.

\clearpage